\documentclass[11pt]{article}
\usepackage{epsfig,natbib,amsmath,float,color,fullpage,amssymb,rotating,bbm}%,pstricks
\usepackage{array}
\newcolumntype{C}[1]{>{\centering\let\newline\\\arraybackslash\hspace{0pt}}m{#1}}
\newcolumntype{L}[1]{>{\raggedright\let\newline\\\arraybackslash\hspace{0pt}}m{#1}}
\newcolumntype{R}[1]{>{\raggedleft\let\newline\\\arraybackslash\hspace{0pt}}m{#1}}

\def\one{\mathbf{1}}
\def\zero{\mathbf{0}}

\def\NN{\mathcal{N}}
\def\CN{\mathcal{CN}}
\def\UU{\mathcal{U}}
\def\IG{\mathcal{IG}}

 \def\b{\mathbf{b}} \def\c{\mathbf{c}}

\def\p{\mathbf{p}}  \def\r{\mathbf{r}}
\def\s{\mathbf{s}}  \def\u{\mathbf{u}}
\def\w{\mathbf{w}} \def\x{\mathbf{x}}
 \def\z{\mathbf{z}}

\def\A{\mathbf{A}} \def\B{\mathbf{B}} \newcommand{\C}{\mathbf{C}}
\def\D{\mathbf{D}}  \def\F{\mathbf{F}}
  \def\I{\mathbf{I}}
 \def\K{\mathbf{K}} \def\L{\mathbf{L}}
\def\M{\mathbf{M}}

\def\V{\mathbf{V}}  
 \def\Z{\mathbf{Z}}

\newcommand{\ep}{\mbox{\boldmath $\epsilon$}}

\newcommand{\et}{\mbox{\boldmath $\eta$}}
\newcommand{\bth}{\mbox{\boldmath $\theta$}}

\newcommand{\la}{\mbox{\boldmath $\lambda$}}
\newcommand{\bmu}{\mbox{\boldmath $\mu$}}

   \def\Th{\mathbf{\Theta}}  
\def\La{\mathbf{\Lambda}}  \def\Si{\mathbf{\Sigma}}

\newenvironment{my_enumerate}{
\begin{enumerate}
  \setlength{\itemsep}{1pt}
  \setlength{\parskip}{0pt}
  \setlength{\parsep}{0pt}}{\end{enumerate}
}

\begin{document}

%\title{Bayesian and Maximum Likelihood Covariance Estimation for Large Gridded Spatial Datasets}
%\title{Parameter Estimation for Stationary Gaussian Processes\\Observed on an Incomplete Lattice}
\title{Bayesian and Maximum Likelihood Estimation for Gaussian Processes on an Incomplete Lattice}

\author{Jonathan R. Stroud, Michael L. Stein and Shaun Lysen\thanks{Stroud is Associate Professor, Department of Statistics, George Washington University (stroud@gwu.edu). 
Stein is Ralph and Mary Otis Isham Professor, Department of Statistics, University of Chicago (stein@galton.uchicago.edu). 
Lysen is Statistician, Quantitative Marketing, Google, Inc. (slysen@google.com).
The work of MLS was supported by the United States Department of Energy, under contract DE-SC0011097.}
\\George Washington University, University of Chicago, and Google, Inc.}

\maketitle

\begin{abstract}

This paper proposes a new approach for Bayesian and maximum likelihood parameter estimation 
for stationary Gaussian processes observed on a large lattice with missing values.  
We propose an MCMC approach for Bayesian inference, and a Monte Carlo EM algorithm for maximum 
likelihood inference.   Our approach uses data augmentation and circulant embedding of the 
covariance matrix, and provides exact inference for the parameters and the missing data.  
Using simulated data and an application to satellite sea surface temperatures in the Pacific Ocean,
we show that our method provides accurate inference on lattices of sizes up to $512 \times 512$, 
and outperforms two popular methods: composite likelihood and spectral approximations.
\end{abstract}

\vfill

%%%%%%%%%%%%%%%%%%%%%%%%%%%%%%%%%%%%%%%%%%%%%%%%%%%%%%%%%%%%%%%%%%%%%%%%%%
\section{Introduction}  
\label{sec:intro}

Spatial lattice data are common in many fields, including environmental 
science, medical imaging and computer modeling.  In these applications, a common
approach is to treat the data as a realization of a stationary Gaussian process, 
and estimate the mean and covariance parameters using maximum likelihood or 
Bayesian methods.  However, in practice the datasets are often extremely large 
and may have missing values.  This makes likelihood inference impracticable, 
since exact Cholesky decompositions require $O(n^3)$ operations, where $n$ is the 
number of observations.  When the observations are taken on a two-dimensional 
lattice and the process is stationary, the exact likelihood cost is reduced to 
$O(n^{5/2})$ \citep{Zimm:89}.  However, when the lattice is incomplete or has 
irregular boundaries, the computational cost is cubic in the number of observations.

To deal with this problem, many approximate likelihood methods have been proposed for 
large spatial datasets.  \cite{Whit:54} introduced a spectral approximation
for lattice data, which has been widely used. \cite{Fuen:07} extended the Whittle approximation
to lattice data with missing values.  \cite{Vecc:88} developed a composite likelihood 
method for unequally-spaced data, and \cite*{SteiChiWelt:04} 
extended this approach to restricted maximum likelihood and provided asymptotic 
standard errors for the parameters.
\cite*{KaufScheNych:08} proposed covariance tapering
for likelihood estimation with unequally-spaced data.   Other 
approaches for large datasets include Markov Random fields \citep*{RueTjel:02}, 
fixed-rank kriging \citep*{CresJoha:08}, predictive processes \citep*{BaneFinlWald:08}, and
predictive processes with tapering \citep*{SangHuan:12}.
However, all of these methods are approximate, so there remains a need for 
exact likelihood-based methods for large gridded datasets.

Recently, \cite*{SteiChenAnit:13} proposed a stochastic method for unbiased
estimation of the score function.  The estimate converges to the true score function as the 
Monte Carlo sample size increases.  However, at present there is no feasible `exact' 
Bayesian Markov chain Monte Carlo solution for this problem, i.e., one 
that converges to samples from the correct posterior distribution as the 
number of iterations increases.

In this paper, we propose a new maximum likelihood and Bayesian approach for
spatial data observed on a large, possibly incomplete, lattice.  The key idea is to view the observed
data as a partial realization from a Gaussian random field on a periodic lattice.  
We then treat the values at the unobserved locations on the periodic lattice
as missing data, and impute them within a data augmentation procedure.  
Conditional on the parameters, the missing data are generated using 
conditional simulation techniques from the geostatistics literature.  Conditional
on the imputed data, we have a complete realization of a periodic process, and the 
complete-data likelihood can be computed efficiently using the fast Fourier 
transform.  This iterative procedure is implemented for Bayesian inference 
using a Markov chain Monte Carlo (MCMC) algorithm, and for maximum likelihood 
estimation using a Monte Carlo expectation-maximization (EM) approach.
Our approach is the first feasible exact MCMC for this setting.

We first use simulated data to show that the methods work well in practice,
and compare them to existing methods.  Under a range of sampling designs,
including complete lattice, missing at random, and missing in blocks, we 
find that the Bayesian approach provides accurate, full probabilistic 
inference for the parameters on lattices up to size 512 $\times$ 512 (262,144 observations).  
Furthermore, our maximum likelihood approach outperforms both composite 
likelihood and spectral approximations in terms of recovering the true 
maximum likelihood estimate.   Finally, we apply the MCMC method to 
a satellite image of sea surface temperatures, where observations are
unavailable over land locations.  The method is shown to provide accurate 
inference in this real-data application.

The rest of the paper is outlined as follows.  In Section 2, we introduce 
the Gaussian process model for lattice data and describe the circulant 
embedding approach. Section 3 provides a MCMC method for Bayesian estimation
and an EM algorithm for maximum likelihood estimation.  The methods are 
illustrated in Section 4 with an extensive simulation study and an
analysis satellite sea surface temperatures.  Conclusions are given in Section 5.  

\section{Likelihood for Gaussian Processes}

Let $\{Z(\s),\s \in D \subseteq \mathbb{R}^d\}$ be a stationary, isotropic 
Gaussian process with mean $\mu$ and covariance function $\mbox{cov}(Z(\s),Z(\s')) 
= \sigma^2 \varphi(|\s-\s'|;\bth)$, where $\varphi(\cdot)$ is an isotropic correlation
function, $|\cdot|$ is Euclidean distance and $\bth$ is a vector of unknown 
parameters.  The goal is to estimate the parameters $(\mu, \sigma^2, \bth)$
based on a realization of the process $\Z = (Z(\s_1), \ldots, Z(\s_n))'$
at the locations $\s_1,\ldots,\s_n$.   The likelihood function is 
\begin{equation}
p(\Z|\mu,\sigma^2,\bth) \; = \; (2\pi\sigma^2)^{-\frac{n}{2}} |\Si(\bth)|^{-\frac{1}{2}}
\exp\left\{-\frac{1}{2\sigma^2}(\Z-\bmu)'\Si(\bth)^{-1}(\Z-\bmu)\right\},
\end{equation}
where $\bmu=\mu\one$, $\one=(1,\ldots,1)'$ and $\Si(\bth)$ is the $n \times n$ 
correlation matrix with elements $\Si_{ij}(\bth) =\varphi(|\s_i-\s_j|;\bth)$.  
If the sampling locations are unequally spaced, the likelihood becomes
computationally infeasible when $n$ is large (say, more than a few thousand), 
because the determinant requires $O(n^3)$ operations to compute.  
If the observations are on a rectangular lattice with no missing data, 
then $\Si(\bth)$ is block Toeplitz with Toeplitz blocks, 
which reduces the cost of the likelihood to $O(n^{5/2})$ \citep{Zimm:89}.  
However, if the observations are on an incomplete lattice (i.e., with 
missing data or non-rectangular boundaries), then $\Si(\bth)$ has no special 
form, and the exact likelihood requires $O(n^3)$ operations.

In this paper, we assume that the data are observed are on a $2-$dimensional lattice 
of size $n_1 \times n_2$. We allow for missing data or irregular boundaries, so the 
lattice may be incomplete.  To implement our estimation approach, we embed the domain 
in a larger lattice of size $N_1 \times N_2$, where $N_1=2rn_1$ and $N_2=2rn_2$ 
and $r\ge 1$.  We consider the periodic extension of $\varphi(h)$ in two dimensions
with period $2r$ in each coordinate.  
Consider the random vector of length $N=N_1N_2$ defined over the embedding lattice.
The covariance matrix of this random vector is block circulant with circulant blocks, 
which allows it to be diagonalized in $O(N\log N)$ operations.  This leads to a 
data augmentation approach where we impute the random field at the unobserved 
locations on the embedding lattice, and then compute the complete data likelihood 
efficiently in $O(N\log N)$ operations.  The approach is detailed below.

\subsection{Circulant Embedding}

Circulant embedding was proposed by \cite{WoodChan:94} and \cite{DietNews:97} 
as a method for simulating stationary Gaussian random fields on a large lattice.
The main idea of this approach is to embed the original $n_1\times n_2$ grid in
$[0,s]^2$ in a larger lattice of size $N_1 \times N_2$, in $[0,2rs]^2$, 
where $r\ge1$, $N_1=2rn_1$, $N_2=2rn_2$, and $N_1$ and $N_2$ are highly 
composite numbers.  We assume throughout the paper that $s=1/\sqrt{2} \approx 0.707$.
Following the notation in \cite{Stei:02} and \cite*{GneiSevcPerc:06}, we define $P_s\varphi$ as the
function on $\mathbb{R}^2$ that has period $2s$ in each coordinate, such that 
$P_s\varphi(\s)=\varphi(|\s|)$, for $\s\in[-s,s]^2$, and let $\C$ denote the 
$N\times N$ covariance matrix obtained by evaluating $P_s\varphi$
over the $N=N_1N_2$ points on the embedding lattice ordered lexicographically. 
Since $P_s\varphi$ is periodic and the domain is a rectangular grid, $\C$ is 
block circulant matrix with circulant blocks (BCCB).  This allows the matrix
to be diagonalized in $O(N\log N)$ operations using the fast Fourier transform (FFT).

To simulate $\Z \sim \NN(\zero,\C)$, we first compute the eigenvalues of $\C$, 
$\la=(\lambda_1,\ldots,\lambda_N)$, then generate an independent random vector 
with variances proportional to the eigenvalues, and then apply an FFT to the 
random vector to obtain a Gaussian random field over the embedding grid.  
The issue is how to choose the value of $r$. The standard embedding approach 
is to choose the smallest value of $r>1$ for which $N_1=2rn_1$ and $N_2=2rn_2$ 
are highly composite numbers.  For some values of $r$, however, this may result
in a non positive-definite matrix $\C$.
To avoid this problem, \cite{WoodChan:94} proposed increasing the value of $r$ until 
$\C$ is positive definite; however, this often requires a very large value 
of $r$, which makes computation prohibitive.

\cite{Stei:02} proposed an alternative approach to ensure positive definite embeddings. 
\cite*{GneiSevcPerc:06} labeled this method {\em cutoff embedding} and explored the 
limits of when the method can be used.  For a given isotropic correlation function 
$\varphi(\cdot)$, \cite{Stei:02} considers the modified correlation function 
$$
\rho(h) = 
\left\{ 
\begin{tabular}{ll} 
$\varphi(h)$   & if $0\le h<1$; \\ 
$\psi(h)$      & if $1\le h < r$; \\
$0$            & if $h \ge r$, 
\end{tabular} 
\right.
$$
where $h$ is Euclidean distance, and $r>1$ is the `cutoff radius' and $\psi(h)$ is a 
function chosen to make $\rho(h)$ differentiable at $r$.  
The compact support of $\rho(\cdot)$ ensures that the periodic function 
$P_r\rho$ is positive definite, providing that $\varphi(\cdot)$ satisfies certain 
conditions \citep{GneiSevcPerc:06}.  \cite{Stei:02} and \cite{GneiSevcPerc:06} 
chose $\psi(h)$ to be a quadratic or square root function.  The circulant embedding 
approach is then applied using the modified covariance function $P_r\rho$, and the 
resulting covariance matrix $\C$ is guaranteed to be positive definite.

\begin{figure}[tbp]
\begin{center}
\includegraphics[width=7cm]{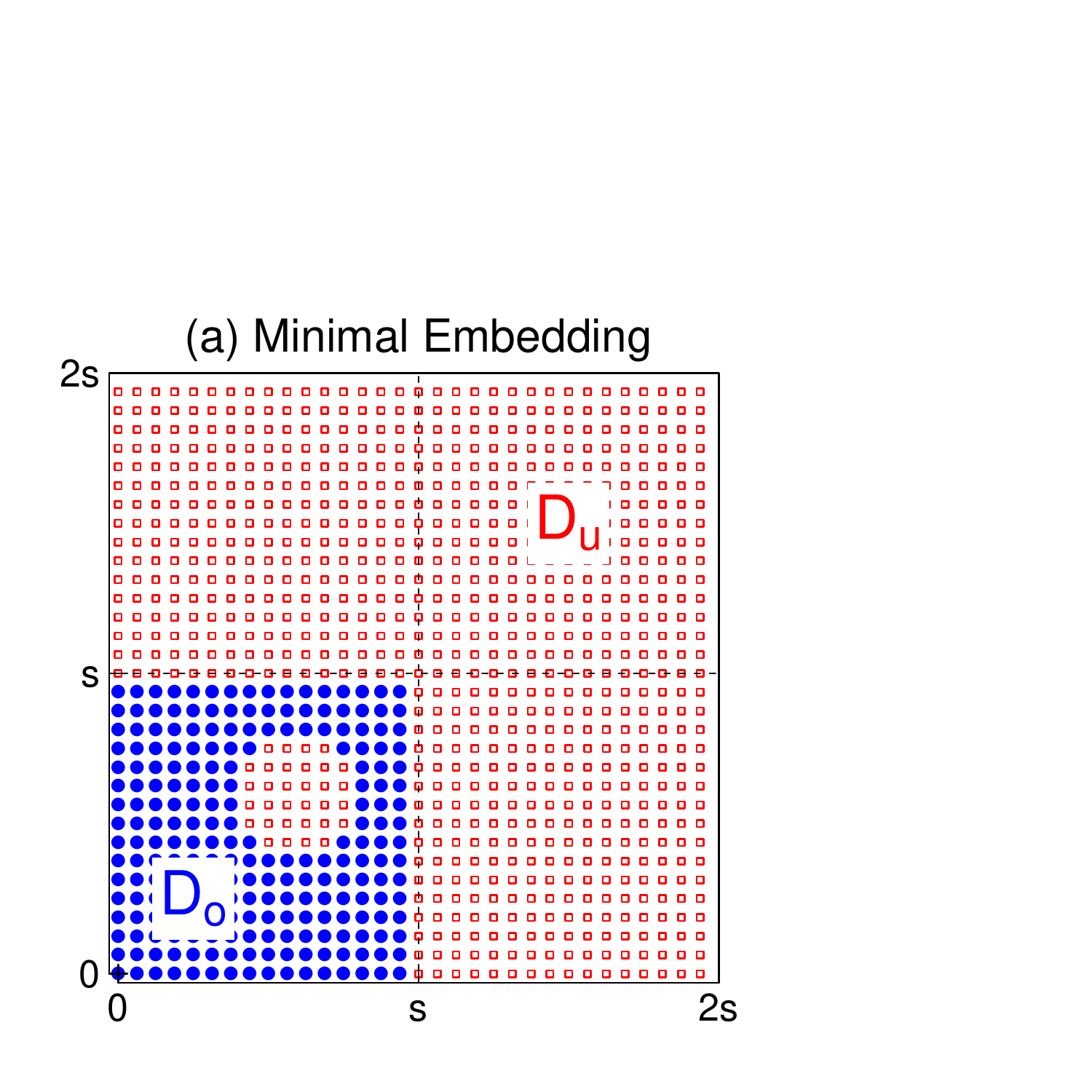}
\includegraphics[width=7cm]{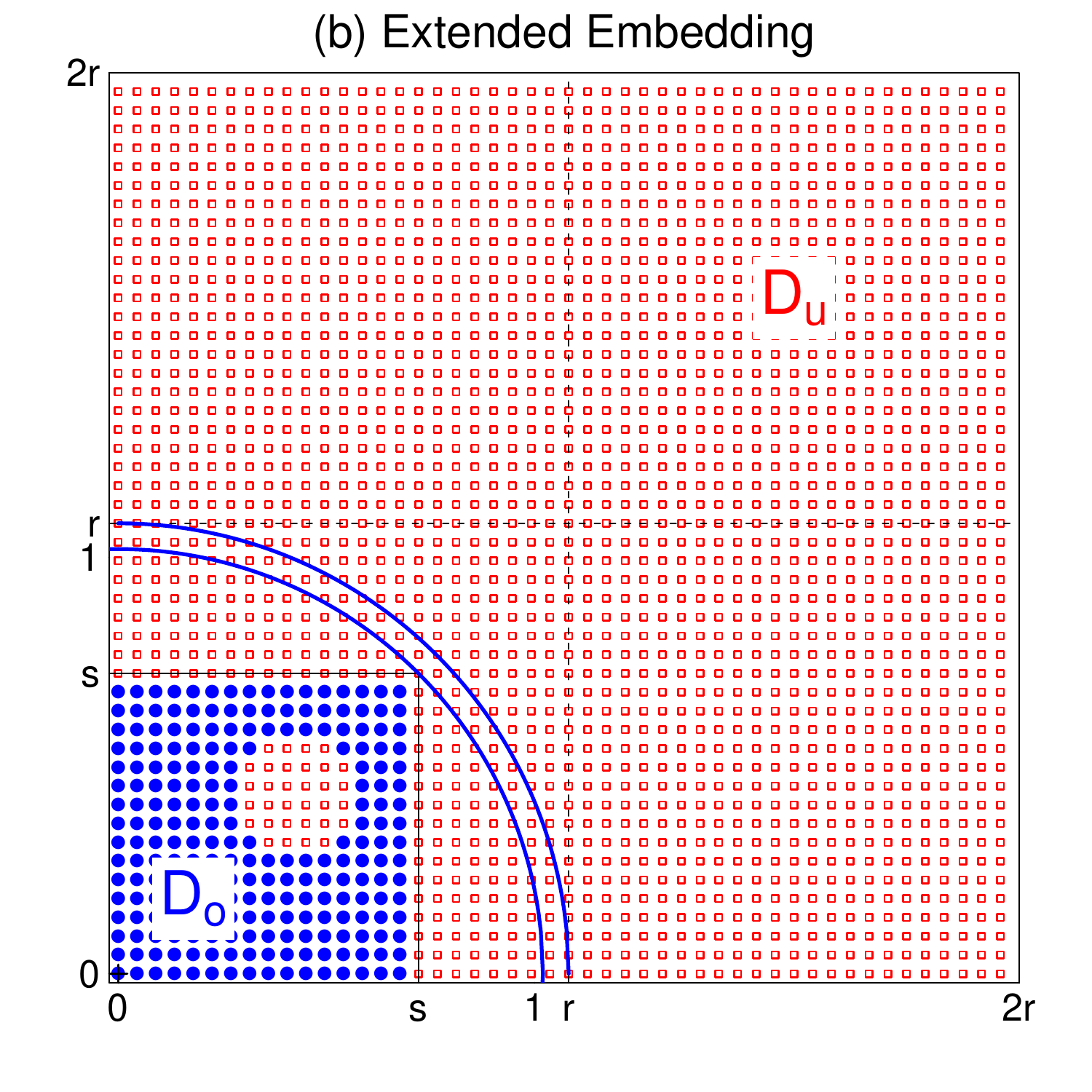}
\caption{Circulant embedding schemes for a square lattice.
Closed circles denote observed locations; open squares denote unobserved locations.
The original domain is $16 \times 16$ on $[0,s]^2$ where $s=1/\sqrt{2} \approx 0.707$,
with data missing in a disk shape.  
(a) Minimal embedding scheme with a $32 \times 32$ embedding lattice.  
(b) Extended embedding scheme with $r=1.5s\approx 1.06$ and a $48\times 48$ embedding lattice.}
\label{fig:embed}
\end{center}
\end{figure}

Figure \ref{fig:embed} illustrates the minimal embedding and extended 
embedding schemes for a square lattice with missing observations.  
Here, the original lattice is $16 \times 16$, with data missing in a circle or 
disk shape.  Panel (a) shows the minimal embedding scheme, where the embedding lattice
is of size $32 \times 32$. Panel (b) illustrates an extended embedding scheme with 
a `cutoff' radius of $r=1.5s\approx 1.06$ and a embedding grid of size $48 \times 48$.

\subsection{BCCB Matrices}

If $\C$ is the covariance matrix for a periodic, stationary random field on a 
$2-$dimensional $N_1\times N_2$ complete lattice with points ordered 
lexicographically, then it has a block circulant form with circulant blocks (BCCB). 
Then, $\C=\F\La\F^*$, where $\F$ is the $2-$dimensional Fourier 
transform matrix, $\F^*$ is the corresponding inverse Fourier transform matrix, 
and $\La= \mbox{diag}(\lambda_1,\ldots,\lambda_N)$ is the diagonal matrix of eigenvalues.
BCCB matrices have a number of computational advantages, namely that eigenvalues, 
matrix-vector multiplications and quadratic forms can be computed efficiently 
in $O(N \log N)$ operations by exploiting the FFT, and they have a storage cost 
of $O(N)$.  The properties of BCCB matrices are summarized in Appendix A.
Also see \cite{Kozi:99} for an excellent summary of BCCB matrices for Gaussian random fields.

%, requiring storage of only the first column of the matrix

\subsection{Unconditional Simulation}
\label{sec:uncondsim}

Exact simulation of periodic, stationary Gaussian random fields on a grid can be 
performed efficiently using the circulant embedding approach of \cite{WoodChan:94}, 
\cite{DietNews:97} and \cite{Stei:02}.  Because the covariance matrix $\C$ is BCCB, 
unconditional simulations can be obtained in $O(N \log N)$ operations by exploiting the 
fast Fourier transform.  Specifically, to generate draws $\Z \sim \NN(\zero,\C)$, 
we set $\Z = \F\La^{1/2}\ep$, where $\ep \sim \CN(\zero,\I)$ is a complex normal random
vector, generated as $\ep=\ep_1+i\ep_2$, with $\ep_1,\ep_2 
\sim \NN(\zero,\I)$.   The vector $\Z=\Z_1+i\Z_2$, yields two independent 
draws $\Z_1$ and $\Z_2$ from $\NN(\zero,\C)$.

\subsection{Likelihood Function}
Let $\Z$ be a random vector representing a stationary, periodic random field on a lattice, 
then it has distribution $\Z \sim \NN(\bmu,\sigma^2\C(\bth))$, where $\C(\bth)$ is BCCB.  
Let $\Th=(\mu,\sigma^2,\bth)$ denote the set of unknown mean and covariance parameters.
The loglikelihood function for the complete data is (ignoring constants)
\begin{eqnarray*}
\log p(\Z|\Th) 
&=& -\frac{N}{2} \log \sigma^2 - \frac{1}{2} \log |\C(\bth)| - \frac{1}{2\sigma^2}(\Z-\bmu)'\C(\bth)^{-1}(\Z-\bmu),\\
&=& -\frac{N}{2} \log \sigma^2 - \frac{1}{2} \sum_{i=1}^N \log \lambda_i - \frac{1}{2\sigma^2}
\left(\La^{-1/2}\F^*\ep\right)'\left(\La^{-1/2}\F^*\ep\right)
\end{eqnarray*}
where $\ep=\Z-\bmu$.  The loglikelihood can be computed efficiently using fast Fourier 
transforms.  We first compute the eigenvalues of $\C(\bth)$ using a 2-dimensional FFT.  
The determinant is then computed as the product of eigenvalues.  The quadratic 
form is computed using an FFT followed by a vector-vector multiplication.  Therefore,
the overall cost to compute the complete-data loglikelihood is $O(N \log N)$ operations.

\subsection{Conditional Simulation}
\label{sec:condsim}
%(i.e., draws of the missing data given the observed data and the parameters).
Our estimation approach requires an efficient method for generating 
conditional simulations of 
the missing data given the observed data and the parameters.
Let $D_o$ and $D_u$ denote the observed and unobserved locations on the embedding lattice, and 
$D=D_o\cup D_u$ denote all locations on the embedding lattice.
Let $\Z_{o}=\{Z(\s): \s \in D_o\}$ and $\Z_{u}=\{Z(\s): \s \in D_u\}$ 
denote the observed and unobserved data on the embedding lattice. 
and let $\Z=\{Z(\s): \s \in D\}$ denote the complete data.  
Suppose $\Z \sim \NN(\bmu,\C)$ and can be partitioned as
\begin{equation}
\left(\begin{tabular}{c} $\Z_o$ \\ $\Z_u$ \end{tabular}\right) 
\sim \NN
\left(
\left(\begin{tabular}{c} $\bmu_o$ \\ $\bmu_u$ \end{tabular}\right) ,
\left(\begin{tabular}{cc} $\C_{oo}$ & $\C_{ou}$ \\ $\C_{uo}$ & $\C_{uu}$ \end{tabular}\right) 
\right)
\label{eqn:mvnormal}
\end{equation}
where $n$ is the number of observed data, $N-n$ the number of unobserved data,
and $N$ is the total number of lattice points. The conditional 
distribution for the missing data given the observed data is
\begin{equation}
\Z_{u} |\Z_{o} \sim \NN\left(\bmu_u + \C_{uo}\C_{oo}^{-1}(\Z_{o}-\bmu_o),\C_{u|o}\right),
\label{eqn:conddist}
\end{equation}
where $\C_{u|o}=\C_{uu}-\C_{uo}\C_{oo}^{-1}\C_{ou}$.    Direct simulation from this 
distribution is infeasible when $N$ is large, because of the cost of computing and storing
the conditional covariance matrix $\C_{u|o}$, and its Cholesky decomposition,
which require $O(N^3)$ and $O(N^2)$ operations, respectively.

We generate conditional simulations using the substitution sampling approach of \cite{Math:76}. 
This method is more efficient than direct simulation, because it avoids the 
conditional covariance matrix and its Cholesky decomposition.   
The method proceeds in two steps.  We first simulate the complete random 
field from its unconditional distribution, $\widetilde{\Z}  \sim 
\NN(\bmu,\C)$, using the approach described in Section~\ref{sec:uncondsim}.   
We then obtain the conditional simulation $\Z^*_u$ by defining
\begin{equation}
\Z^*_{u}= \widetilde{\Z}_{u} + \C_{uo}\C_{oo}^{-1}(\Z_{o}-\widetilde{\Z}_{o}).
\label{eqn:condsim}
\end{equation}
It is straightfoward to show that $\Z^*_{u} \sim p(\Z_{u}|\Z_{o},\Th)$;
i.e., it has the mean and covariance given in (\ref{eqn:conddist}).  
For a proof, see \cite{ChilDelf:12}.  Note that $\Z^*_u$ has the same 
form as the conditional mean given in (\ref{eqn:conddist}), but with the 
simulated field $\widetilde{\Z}$ substituted for $\bmu$.   
To obtain the conditional simulation, we must first solve the system
\begin{equation}
\C_{oo}\x=\et,  \mbox{ where }   \et=\Z_o-\widetilde{\Z}_o.
\label{eqn:linsys}
\end{equation}
It is infeasible to solve this directly when $n$ is large, as it requires $O(n^3)$ 
operations.  Instead we use an iterative method to solve the system, which is described below.  
After solving the system, the conditional simulation is obtained by computing $\w_u=\C_{uo}\x$,
which can be done efficiently by exploiting the form of $\C$,  
and then setting $\Z^*_{u}= \widetilde{\Z}_{u} + \w_u$.

\subsection{Preconditioned Conjugate Gradient}

We use the preconditioned conjugate gradient (PCG) algorithm \citep{GoluVanl:96} to
solve the system (\ref{eqn:linsys}).  The PCG is an iterative method that solves the 
modified system 
\begin{equation}
\M^{-1}\C_{oo}\x=\M^{-1}\et 
~~ \mbox{where}~~  \et=\Z_o-\widetilde{\Z}_o,
\label{eqn:linsys-pcg}
\end{equation}
where $\M^{-1}$ is an $n\times n$ preconditioner matrix.
The solution to the modified system is the same as the original solution, but the 
use of the preconditioner speeds up convergence of the algorithm (see Appendix B).  
Convergence to the exact solution is guaranteed within $n$ iterations; 
however, a good approximation can usually be obtained in far fewer iterations.  
The algorithm is stopped at the iteration $k$ when the residual vector 
$\r_k=\et-\C_{oo}\x_k$, is smaller than a specified tolerance.  We use the 
criterion $|\r_k|/|\r_0|<\epsilon$, where $\epsilon$ is a specified error tolerance.

The PCG requires only matrix-vector multiplications 
of the form $\C_{oo}\x$ and $\M^{-1}\x$.  The former can be computed efficiently 
by exploiting the block circulant structure of $\C$.  Suppose $\C$ is partitioned
as in (\ref{eqn:mvnormal}).  To compute $\C_{oo}\x$, we pad the vector $\x$ with zeros, 
i.e., $\x^*=(\x',\zero')'$, then multiply $\w=\C\x^*$, and the result is obtained in 
the first $n$ elements of $\w$:
$$
\w=\C\x^* = 
\left(
\begin{tabular}{cc}
$\C_{oo}$ & $\C_{ou}$ \\
$\C_{uo}$ & $\C_{uu}$ 
\end{tabular}
\right)
\left(
\begin{tabular}{c}
$\x$ \\ $\zero$
\end{tabular}
\right)
=
\left(
\begin{tabular}{c} $\C_{oo}\x$ \\ $\C_{uo}\x$ \end{tabular}
\right)
$$
This procedure is also used to compute $\C_{uo}\x$, but the result
is obtained in the last $N-n$ elements of $\w$.  This step is needed in 
the conditional simulations after solving the system.  Each multiplication of the form
$\C\x$ requires two fast Fourier transforms, which require $N\log N$ operations.  
Each PCG iteration requires one $\C\x$ multiplication.  Therefore, the total computational 
cost for one conditional simulation is $O(IN\log N)$, where $I$ is the number of PCG 
iterations.

\subsection{Preconditioners}

%A good choice of preconditioner is crucial for the performance of the PCG.
The performance of the PCG depends critically on the choice of preconditioner.
Ideally, the preconditioner should satisfy three criteria: $\M^{-1}\C_{oo}$ should 
have a small condition number; $\M^{-1}\x$ can be multiplied quickly;
and $\M^{-1}$ should have a low storage cost.   Common choices for preconditioners 
include circulant/block circulant matrices, block diagonal matrices, incomplete LU 
or Cholesky decompositions, and sparse matrices \cite[see][]{GoluVanl:96}.

In our analysis, we propose a new preconditioner based on the 
composite likelihood methods of \cite{Vecc:88} and \cite{SteiChiWelt:04}.  
For this method, the observed data $\Z$ are partitioned into $q$ blocks and the likelihood 
is approximated by a product of conditional normal densities $p(\A_j\Z|\B_{j}\Z)$, 
$j=1,\ldots,q$, where $\A_j$ and $\B_j$ are matrices of zeros and ones that define 
the prediction and conditioning sets for block $j$.  The sets are chosen to be small 
so that the conditional moments can be computed and stored efficiently.  
Since each conditional density is Gaussian, the likelihood approximation corresponds to a 
multivariate normal density $\NN(\Z|\zero,\V)$, where $\V^{-1}=\L'\D\L$, where $\L$ 
and $\D$ are sparse $n\times n$ matrices containing the regression coefficients and 
precision matrices for the conditional distributions (see Appendix C).  The 
approximation assumes that $\NN(\Z|\zero,\V) \approx \NN(\Z|\zero,\C_{oo})$.
Therefore, we choose $\V^{-1}$ as our preconditioner.  We then only need to
specify the prediction and conditioning sets.  In our appplications, we choose 
prediction sets of size 4 and conditioning sets of size 18, 33 or 52.

We have also developed a number of other preconditioners, including BCCB, block diagonal, 
and sparse covariance/precision matrices based on Whittle's 
approximation, covariance tapering and Markov random fields, respectively.  One preconditioner 
that works quite well is the observed block of the complete-data precision matrix, i.e., $(\C^{-1})_{oo}$.  
Since the inverse of a BCCB matrix is also BCCB, it has a storage cost of $O(N)$ and
matrix-vector multiplications are computed efficiently using FFTs.  We found that this 
preconditioner works well for complete or nearly complete lattices, but less well
with large amounts of missing data.
However, we found that all of these choices were generally outperformed by the Vecchia
preconditioner in terms of convergence rate and run time.

In the next section, we propose two estimation algorithms based on the ideas of
circulant embedding and conditional simulation.  First, we propose a 
MCMC algorithm for Bayesian inference.  Second, we introduce a Monte Carlo 
EM algorithm for maximum likelihood estimation.   We emphasize
that both of the proposed algorithms are {\em exact}, up to Monte Carlo error.

\section{Parameter Estimation}

\subsection{Bayesian Estimation}
\label{sec:Bayes}

For the Bayesian analysis, we specify a prior distribution for the 
unknown parameters, $\pi(\Th)$, and make inference based on the joint 
posterior distribution 
$$
\pi(\Th,\Z_{u}|\Z_{o}) \propto p(\Z|\Th) \, \pi(\Th),
$$
where $\Z=(\Z_{o},\Z_{u})$ denotes the complete data.   This joint posterior
distribution is typically unavailable in closed form.  Therefore, we propose a Markov 
chain Monte Carlo (MCMC) algorithm to sample from it.
Specifically, we propose a two-block Gibbs sampler that alternates between 
updating the missing data and the parameters.  Given initial parameter values,
$\Th^0$, the MCMC algorithm proceeds as follows for $i=1,\ldots,M$:
\begin{my_enumerate}
\item Generate $\Z_u^i \sim p(\Z_u|\Z_o,\Th^{i-1})$.
\item Generate $\Th^i \sim \pi(\Th|\Z^{i})$.
\end{my_enumerate}
The missing data are updated using 
conditional simulation methods described in Section~\ref{sec:condsim}.   
The parameters are updated using a block Metropolis-Hastings 
step, described below.

Given the complete data, the parameters are generated from their conditional 
distribution $\pi(\Th|\Z) \propto p(\Z|\Th) \pi(\Th)$.  This distribution is 
easy to evaluate, but is generally unavailable in closed form due to
the nonlinearity of $\bth$ in the determinant and the quadratic form.  Bayesian 
MCMC approaches to estimate covariance parameters in spatial Gaussian processes 
include \cite{EckeGelf:97}, who proposed a Metropolis algorithm and \cite{AgarGelf:05}, 
who proposed a slice sampling approach.  Here, we use a Metropolis-Hastings 
scheme to update the parameters, which proceeds as follows.
\begin{my_enumerate}
\item Generate $\Th^{*}$ from a proposal distribution $q(\Th|\Th^{i})$.  
\item Accept $\Th^{*}$ with probability 
$$
\min \left\{1,  
     \frac{p(\Z|\Th^{*})\,\pi(\Th^{*})}{p(\Z|\Th^{i})\, \pi(\Th^{i})}
     \frac{q(\Th^{i}|\Th^{*})}{q(\Th^{*}|\Th^{i})}
     \right\}.  
$$
\end{my_enumerate}

The computational cost to generate the missing data is $O(IN\log N)$, and
the cost to update the parameters is $O(N\log N)$.  Hence the cost for each 
iteration of the Gibbs sampler is $O(IN\log N)$ operations, 
and the total cost for $M$ iterations of the sampler is $O(MIN\log N)$.

The Bayesian approach also provides inference for the random field at missing
locations via the posterior predictive distribution \cite[see][]{HandStei:93}.
If the missing data lie on the original lattice or the embedding lattice, 
the MCMC algorithm automatically generates samples from their distribution as
part of the imputation step.

\subsection{Maximum Likelihood Estimation}

For maximum likelihood (ML) estimation, we propose an expectation-maximization (EM) 
algorithm \cite*[][]{DempLairRubi:77} to obtain the maximum likelihood estimate $\widehat{\Th} = 
\arg \max_{\Th} p(\Z_{o}|\Th)$.  The algorithm iterates between the E-step and M-step
until convergence. In the E-step, we calculate the expected complete-data 
loglikelihood given the observed data and the current parameter, $\Th^t$:
\begin{equation}
Q(\Th|\Th^{t}) = \int \log p(\Z|\Th) p(\Z|\Z_o,\Th^{t}) d\Z.
\label{eqn:Q}
\end{equation}  
In the M-step, we maximize this function to obtain the next parameter value, $\Th^{t+1}$. 
Under the Gaussian model (\ref{eqn:mvnormal}), the distribution for the complete data is 
$\Z|\Th \sim \NN(\bmu(\Th),\C(\Th))$, and the conditional distribution for the complete data is 
$\Z|\Z_o,\Th^t \sim \NN(\widetilde\bmu(\Th^t),\widetilde\C(\Th^t))$. 
Suppressing dependence on parameters, and ignoring constants, the expectation (\ref{eqn:Q}) is 
\begin{eqnarray}
Q(\Th|\Th^{t}) &=& -\frac{1}{2} \log |\C| - \frac{1}{2} E\left\{(\Z-\bmu)'\C^{-1}(\Z-\bmu)|\Z_o,\Th^{t}\right\}
\label{eqn:Q1}\\
&=& -\frac{1}{2}\log |\C| -\frac{1}{2}\left\{\mathrm{tr}(\C^{-1}\widetilde\C)+ 
(\widetilde{\bmu}-\bmu)'\C^{-1}(\widetilde{\bmu}-\bmu)\right\}.
\label{eqn:Q2}
\end{eqnarray}
Note that the trace term in (\ref{eqn:Q2}) involves  $\widetilde\C$, the conditional 
covariance matrix for the complete data.  This matrix consists of $\widetilde{\C}_{u|o}$ 
in the lower diagonal block, and zeros elsewhere, and does not have a BCCB form.
Thus, it is infeasible to compute $\widetilde{\C}$ when $n$ is large and therefore 
$Q$ cannot be evaluated exactly.  Instead, we propose a Monte Carlo approach, 
where we approximate the expected loglikelihood (\ref{eqn:Q1}) by
\begin{equation}
\widehat{Q}(\Th|\Th^{t}) = -\frac{1}{2} \log|\C| - \frac{1}{2} 
\left\{\frac{1}{M} \sum_{i=1}^M (\Z^{(i)}-\bmu)'\C^{-1}(\Z^{(i)}-\bmu)\right\},
\label{eqn:Qapprox}
\end{equation}
where $\Z^{(1)},\ldots,\Z^{(M)} \sim p(\Z|\Z_o,\Th^t)$ are conditional simulations
of the complete data generated using the current parameter value $\Th^t$.  
We then maximize $\hat{Q}$ to obtain the new 
parameter value, $\Th^{t+1}$.  The Monte Carlo expectation avoids computing the conditional 
covariance matrix, and requires only a determinant and $M$ quadratic forms involving 
the matrix $\C$.  Therefore, $\hat{Q}$ can be computed for large $n$.
Given an initial parameter $\Th^0$, the EM algorithm proceeds as follows for $t=0,1,\ldots,T$.

\vspace{-.3cm}
\begin{my_enumerate}
\item {\bf (E-step)} Generate $\Z_u^{(1)},\ldots,\Z_u^{(M)} \sim p(\Z_u|\Z_o,\Th^t).$
\item {\bf (M-step)} Update $\Th^{t+1} = \arg \underset{\Th}{\max}\,  \widehat{Q}(\Th|\Th^t)$.   
\end{my_enumerate}
In Step 1 of the EM algorithm,  we generate $M$ conditional simulations using
the current parameter $\Th^t$ using the approach described in Section~\ref{sec:condsim}.
In Step 2, we maximize the expected complete-data loglikelihood using numerical 
optimization methods such as Newton-Raphson or the Nelder-Mead simplex algorithm.
%In some cases it is possible to analytically maximize for some of the parameters,
%using either the true or the approximate $Q$ function.

\section{Examples}

\subsection{Simulation Study}

To study the performance of the estimation algorithms, we first conduct a 
detailed simulation study using different lattice sizes and missingness patterns.  
We generate data from a stationary, isotropic Gaussian process, 
with mean $\mu$ and covariance $\mathrm{Cov}(Z(\s),Z(\s'))=\sigma^2\varphi(|\s-\s'|)$, 
where $\varphi(\cdot)$ is a powered exponential correlation with 
microscale noise
\cite[]{Cres:93}: 
\begin{equation}
\varphi(h) = \exp\left\{ -(h/\lambda)^\alpha \right\} + c \mathbbm{1}_{(h=0)}.
\label{eqn:pow-exp}
\end{equation}
Here $\sigma^2$ is the partial sill parameter, $\lambda>0$ is the spatial range 
parameter, $\alpha \in (0,2]$ is the shape parameter, $c\ge 0$ is the ratio between
the microscale and macroscale variation, and $\mathbbm{1}_A$ is an indicator for the event $A$.  
The parameter $\tau^2 = c\sigma^2$ represents the variance of the microscale noise.
The model contains the exponential ($\alpha=1$) and squared exponential ($\alpha=2$)
covariances as special cases.  For the simulation study, we fix $c$ at its 
true value and estimate the parameters $\mu$, $\sigma^2$ and $\bth=(\lambda,\alpha)$, 
using the methods from Section 3.   This is done to make the simulation study feasible
while still allowing a nugget effect, which is often present in practice.

For circulant embedding, we use a variant of the cutoff embedding 
approach described in Section 2, with the modified correlation function
\begin{equation}
\rho(h) = 
\left\{ 
\begin{tabular}{ll} 
$\exp\{-(h/\lambda)^\alpha\} + c \mathbbm{1}_{(h=0)},$ & $0\le h < 1$; \\
$a + b (h-r)^2,$    & $1 \le h < r$; \\
$a,$                & $h \ge r$,
\end{tabular} 
\right.
\label{eqn:embed}
\end{equation}
where $r>1$, and $a=\exp\{-(1/\lambda)^\alpha\}/\{1-(r-1)/2\lambda\}$ and
$b=\exp\{-(1/\lambda)^\alpha\}/\{2\lambda(r-1)\}$ are chosen to make $\rho(h)$ 
differentiable at $1$ and $\rho'(r)=0$.   This approach is similar to cutoff 
embedding with a quadratic function $\psi(h)$, but here $r$ is selected by 
the user, and $\rho(h)$ is set to a constant rather than zero for $h\ge r$.  
While this approach does not guarantee non-negative definite embeddings, it 
leads to fewer violations than standard embedding, while allowing for a 
much smaller value of $r$ than required for cutoff embedding.  

%for MCMC or EM algorithm,
The value of $r$ required for a non-negative definite embedding depends on 
the parameters $\lambda, \alpha, c$.  If these parameters were known, we could 
choose $r$ by trial and error.  However, in the context of an MCMC or EM algorithm, 
the parameters are unknown and changing at each iteration.  One possible solution 
is to adaptively update $r$ (and the size of the embedding grid)
along with the parameters.   However, this implies a variable-dimensional 
state space, which requires reversible jump or other trans-dimensional MCMC 
methods, which are difficult to use in high-dimensional settings.  To simplify 
estimation, we hold $r$ fixed throughout the estimation algorithm, and choose its
value based on prior information, with subsequent modifications made based on a 
few trial runs of the algorithm.

For the simulation study, we generate data on a $n_1 \times n_1$ lattice
on $[0,s]^2$, where $s=1/\sqrt{2}$.  The true parameter values are $\sigma^2=4$, $\lambda=0.10$, 
$\alpha=1$, $c=0.01$ and $\mu=10$, which corresponds to an exponential covariance with 
small microscale variation.  We use the modified correlation function (\ref{eqn:embed})
with $r=1.5s\approx 1.06$, and an embedding lattice of size $3n_1 \times 3n_1$ on $[0,3s]^2$.  
The `cutoff' radius $r$ is about 10 times larger than the spatial range parameter $\lambda$.
We focus on the behavior of the algorithms over a fixed spatial region as the grid becomes increasingly dense. 
We consider observation lattices of size $n_1=$ 32, 64, 128, 256 and 512 
with corresponding embedding lattices of size $N_1=$ 96, 192, 384, 768, and 1536,
and three different missingness patterns: complete lattice, 
10\% missing at random, and 10\% missing disk.

\subsubsection{Bayesian Analysis}

For the Bayesian analysis, we assume a prior distribution of the form
$\pi(\mu,\sigma^2,\bth)\propto \pi(\bth)/\sigma^2$, 
corresponding to a noninformative Jeffreys' prior for the mean and variance, 
$\pi(\mu,\sigma^2)\propto 1/\sigma^2$, and a prior for the correlation
parameters, $\pi(\bth)=\pi(\lambda)\pi(\alpha)$, where
\begin{equation*}
\pi(\lambda) =  \frac{0.5}{(1+0.5\lambda)^2}
\end{equation*}
and $\pi(\alpha) = \UU(0,2)$.  
Note that a proper prior for the range parameter is needed to ensure a proper 
posterior \citep*{BergDeolSans:01}.  We choose a proper prior for $\lambda$
 with a mode of zero, a median of two, and a long right tail, 
reflecting our belief that large values of the range are less likely than small ones.
A similar prior was used by \cite{HandStei:93} and \cite{HandWall:94}. 
This prior is uninformative for $0.5\lambda/(1+0.5\lambda)$ on $[0,1]$.
The prior for the shape parameter is proper and uniform over its support $\alpha \in (0,2]$.

Let $\Z$ denote the complete data over the embedding lattice.  Then $\Z \sim \NN(\mu\one,
\sigma^2\C(\bth))$, where $\C(\bth)$ is the correlation matrix obtained by evaluating 
$P_r\rho(h)$ over the points on the embedding lattice.  Multiplying the prior 
distribution and the complete-data likelihood, we obain the full conditional 
posterior for the parameters:
$$
\pi(\mu,\sigma^2,\bth|\Z) ~\propto~
 (\sigma^2)^{-\frac{N}{2}-1} |\C(\bth)|^{-\frac{1}{2}} 
%\exp\left\{-\frac{1}{2\sigma^2}\left[(\mu-\hat\mu)\one'\C(\bth)^{-1}\one(\mu-\hat\mu)+S^2(\bth)\right]\right\} 
\exp\left\{-\frac{1}{2\sigma^2}\left[\frac{(\mu-\hat\mu)^2}{(\one'\C(\bth)^{-1}\one)^{-1}}+S^2(\bth)\right]\right\} \, \pi(\bth), 
$$
where $\hat\mu=\one'\C(\bth)^{-1}\Z/(\one'\C(\bth)^{-1}\one) = \one'\Z/N=\bar{Z}$ 
and $S^2(\bth)=(\Z-\hat\mu\one)'\C(\bth)^{-1}(\Z-\hat\mu\one)$ 
are the generalized least squares estimate and sum of squares, respectively.  
Note that because $\C(\bth)$ is a BCCB matrix, the least squares estimate 
$\hat\mu$ does not depend on $\bth$, 
and the determinant and sum of squares can be computed efficiently.

We follow the MCMC approach described in Section~\ref{sec:Bayes}, but improve the
efficiency of the algorithm by updating all parameters as a block. To do this, 
we factorize the full conditional posterior for the parameters as $\pi(\mu, \sigma^2,
\bth\,|\,\Z)$ $= \pi(\mu|\sigma^2,\bth,\Z)$ $\pi(\sigma^2\,|\,\bth,\Z)$ $\pi(\bth\,|\,\Z)$, 
where 
\begin{eqnarray}
\pi(\mu|\sigma^2,\bth,\Z)  & = & \NN(\bar{Z},\sigma^2(\one'\C(\bth)^{-1}\one)^{-1})\nonumber\\
    \pi(\sigma^2|\bth,\Z)  & = & \IG\left((N-1)/2,S^2(\bth)/2\right)\nonumber\\
     \pi(\bth\,|\,\Z) & \propto & |\C(\bth)|^{-\frac12} |\one'\C(\bth)^{-1}\one|^{-\frac12}
\left\{S^2(\bth)\right\}^{-\frac{N-1}{2}} \, \pi(\bth).
\label{eqn:th-post}
\end{eqnarray}
The conditional posterior for $(\mu,\sigma^2)$ has the standard conjugate normal-inverse gamma form.
The marginal posterior for $\bth$ in (\ref{eqn:th-post}) is not of a recognizable form, but
can be efficiently evaluated pointwise, since the determinants and sum of squares involve the
BCCB matrix $\C(\bth)$.  This leads to a Metropolis-Hastings 
algorithm to generate the parameters as a block from their full conditional 
distribution $(\mu,\sigma^2,\bth)\sim \pi(\mu,\sigma^2,\bth|\Z)$.
Given the current values $(\Z^i, \bth^i)$, the parameter update is as follows:
\begin{my_enumerate}
\item Draw a candidate value $\bth^* \sim q(\bth|\bth^i)$.
\item Accept $\bth^*$ with probability
\begin{equation}
\min\left\{1, \frac{\pi(\bth^*|\Z^i)}  {\pi(\bth^i|\Z^i)}
              \frac{q(\bth^i|\bth^*)}{q(\bth^*|\bth^i)} \right\}.
\end{equation}
\item If $\bth^*$ is accepted, draw $\sigma^2 \, \sim \, \IG((N-1)/2,S^2(\bth^*)/2)$,
and $\mu \sim \NN(\bar{\Z^i},\sigma^2(\one'\C(\bth^*)^{-1}\one)^{-1})$;
otherwise, leave $(\mu,\sigma^2)$ unchanged.
\end{my_enumerate}
We choose the proposal distribution $q(\bth|\bth^i)$ to be a bivariate lognormal, 
with covariance matrix chosen to achieve an acceptance probability of around 35\%.
We note that a similar Metropolis algorithm with block updating was proposed by  \cite*{HuerSansStro:04} 
in the context of spatio-temporal models.  They found that blocking provides 
huge gains in computational efficiency relative to updating each parameter one at a time,
particularly for large datasets.

\begin{figure}[tbp]
\begin{center}
\includegraphics[width=15cm]{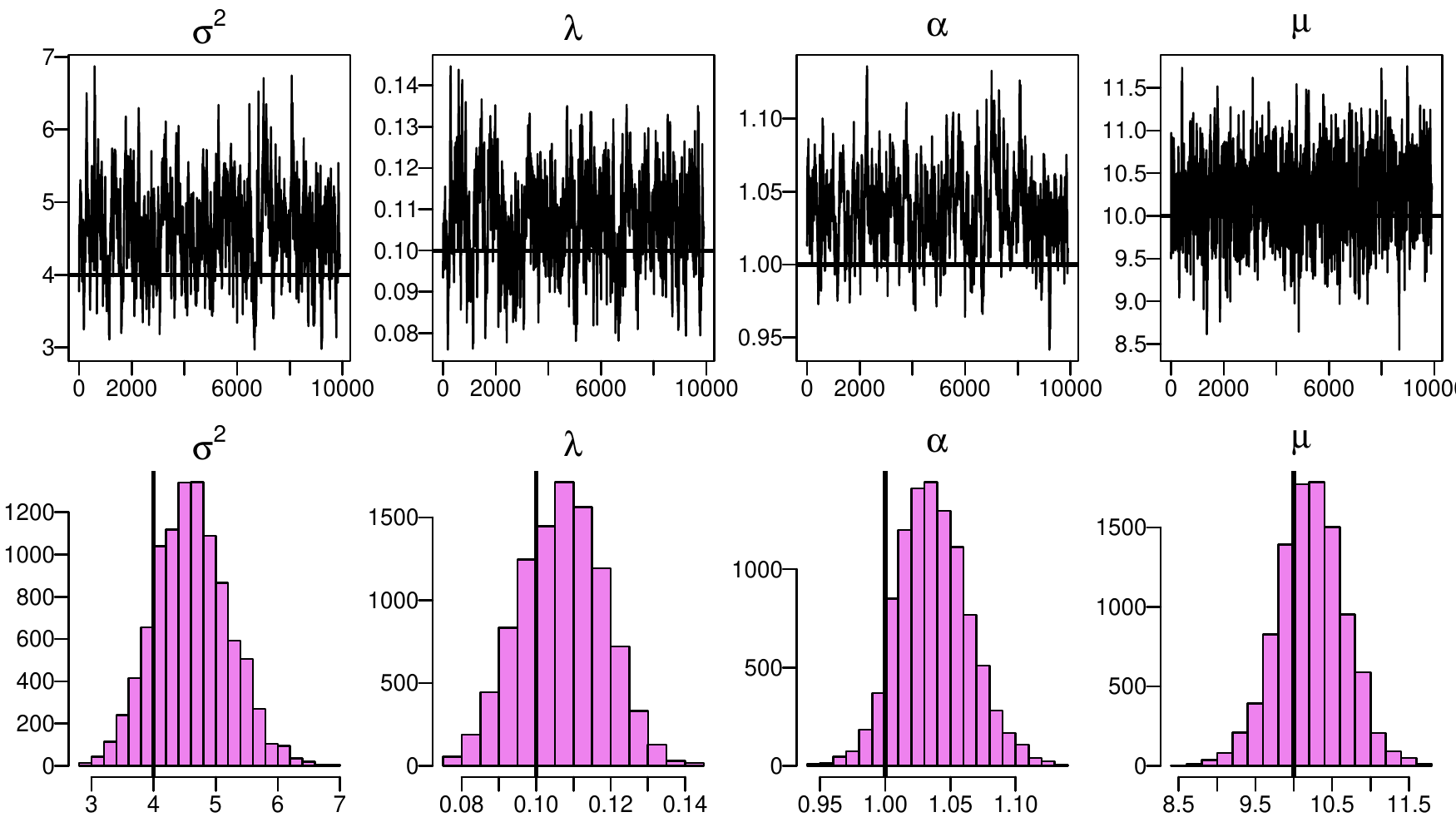}
\caption{MCMC trace plots and histograms for the parameters. Results are based on simulated 
data on a $128 \times 128$ lattice with 10\% missing data in a disk shape.
The true parameter values are $\sigma^2=4$, $\lambda=0.1$, $\alpha=1$ and $\mu=10$.}
\label{fig:mcmc-sim90}
\end{center}
\end{figure}

\begin{figure}[tbp]
\begin{center}
\includegraphics[width=16cm]{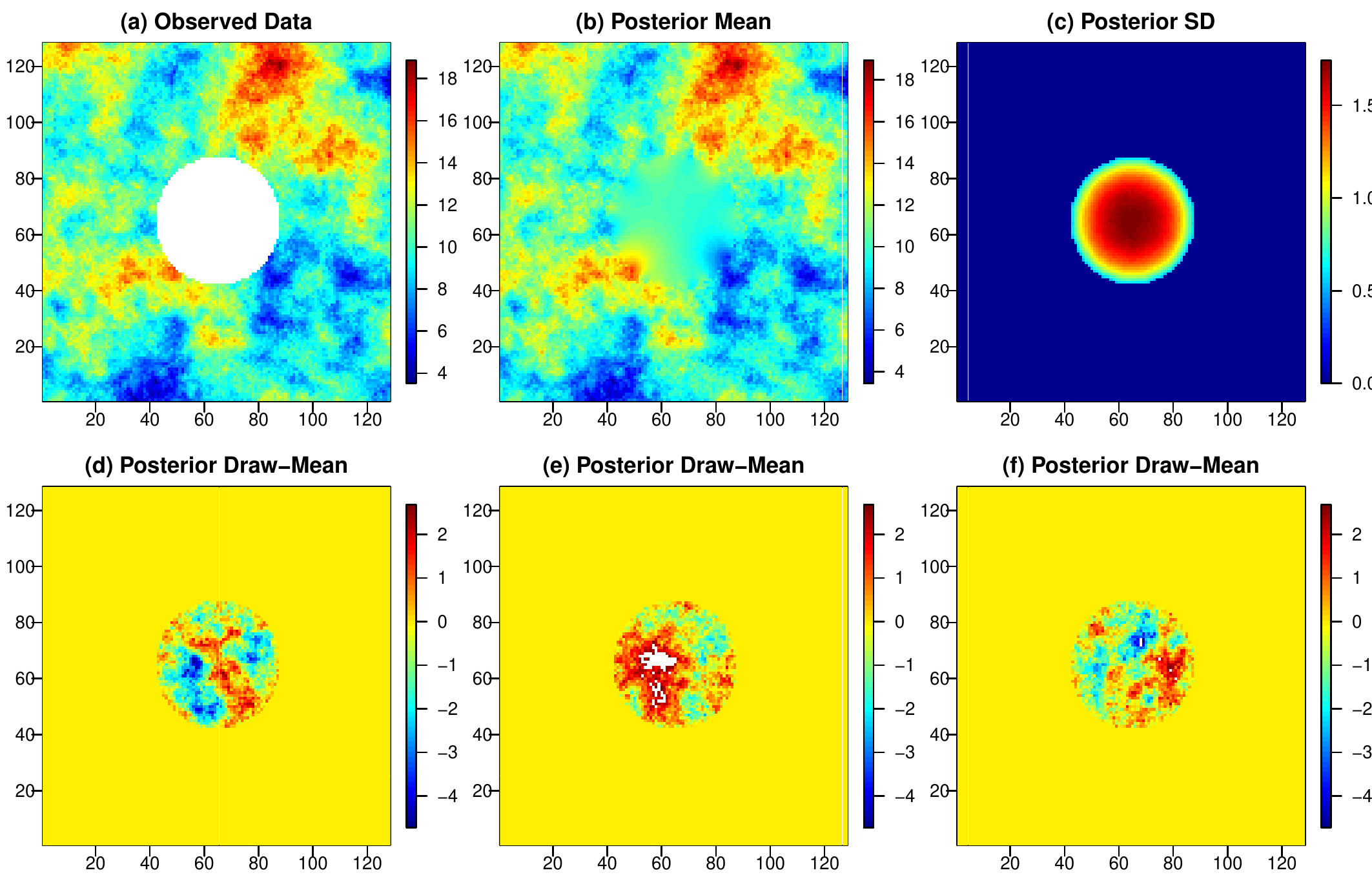}
\caption{MCMC posterior summaries of the random field for simulated data 
on a $128 \times 128$ lattice with 10\% missing data in a disk shape.  
(a) observed data $Z(\s)$.
(b)-(c) posterior mean and standard deviation for $Z(\s)$.
(d)-(f) posterior draws minus the posterior mean for $Z(\s)$.}
\label{fig:mcmc-draws90}
\end{center}
\end{figure}

The results from the MCMC algorithm are reported in Table~\ref{table:MCMC}
and Figures~\ref{fig:mcmc-sim90}--\ref{fig:mcmc-draws90}. All results are based on 
the Vecchia preconditioner with prediction sets of size 4 and conditioning
sets of size 52, with a PCG tolerance of $\epsilon=10^{-5}$.  
Calculations are implemented in C on an Intel Xeon 2.8 GHz processor with 22 GB 
of RAM on a Mac OS X operating system.  Fast Fourier transforms are implemented 
with the FFTW package \citep{FrigJohn:98}.

Figures~\ref{fig:mcmc-sim90} and \ref{fig:mcmc-draws90} show the results from the
MCMC analysis of simulated data on a $128 \times 128$ lattice with 10\% missing 
values in a disk-shaped region.  In total, there are 14,743 observations.  
The MCMC described above was run for 10,000 iterations after a burn-in period of 1000.
Figure~\ref{fig:mcmc-sim90} shows posterior trace plots and histograms for the parameters.
The trace plots are stable, indicating no clear violations of stationarity.  
The histograms are all unimodal and fairly symmetric.  All of the histograms 
contain the true parameter values, and all of the 95\% posterior intervals 
(not shown) contain the true parameter values.  This indicates that the algorithm 
is providing accurate samples from the posterior distribution.

Figure~\ref{fig:mcmc-draws90} summarizes the posterior distribution of the spatial 
field.  Panels (a) shows the observed data $Z(\s)$, (b) and (c) show the posterior 
mean and standard deviation for the field, and panels (d)-(f) show three posterior 
draws minus the posterior mean.  Note that the posterior mean agrees with the 
observed data at the observation locations, and converges to the unconditional mean 
($\hat\mu\approx 10$) near the center of the domain.  In addition, there is no 
posterior uncertainty for $Z(\s)$ at the observation locations, and the posterior 
standard deviation converges to the unconditional SD ($\hat\sigma\approx 2$) near 
the center of the domain.   Panels (d)-(f) illustrate the sample-to-sample 
variation in the posterior field; the plots illustrate the correlation length
scales and reiterate that the uncertainty is highest in the center of the domain.

Next, we study the behavior of the MCMC algorithm for different lattice sizes and 
different missingness patterns.   As stated earlier, we consider increasingly dense 
lattices of size $n_1 \times n_1$ with $n_1=$ 32, 64, 128, 256 and 512, 
with corresponding embedding lattices of size $3n_1 \times 3n_1$.  We consider three 
designs: complete lattice, 10\% missing at random, and 10\% missing disk.  
The results described below are based on 2000 MCMC iterations after a burn-in period 
of 500.  The true parameter values, choice of $r$, and proposal distribution, are 
the same as above.

% latex table generated in R 3.0.1 by xtable 1.7-1 package
% Fri Jan 24 09:12:48 2014
\begin{table}[tbp]
\centering
\begin{tabular}{cccccccc}
  \hline \hline 
    $n_1$& $n$ & $\sigma^2$& $\lambda$& $\alpha$& $\mu$& Iter& Time\\ \hline 
 \multicolumn{8}{c}{{\em Complete Lattice}} \\ 
   32 &   1024 & 4.25 {\small (0.92)} & 0.094 {\small (0.024)} & 1.057 {\small (0.057)} & 9.48 {\small (0.40)} & 5 & 1 \\ 
   64 &   4096 & 3.10 {\small (0.69)} & 0.090 {\small (0.026)} & 0.942 {\small (0.027)} & 9.84 {\small (0.34)} & 8 & 3 \\ 
  128 &  16384 & 4.38 {\small (0.47)} & 0.103 {\small (0.010)} & 1.030 {\small (0.022)} & 10.35 {\small (0.40)} & 13 & 23 \\ 
  256 &  65536 & 4.10 {\small (0.27)} & 0.100 {\small (0.004)} & 1.012 {\small (0.021)} & 9.89 {\small (0.36)} & 22 & 167 \\ 
  512 & 262144 & 4.05 {\small (0.09)} & 0.098 {\small (0.002)} & 1.011 {\small (0.010)} & 9.66 {\small (0.36)} & 38 & 1248 \\ 
   \multicolumn{8}{c}{{\em Missing at Random (10\%)}} \\ 
   32 &   922 & 3.94 {\small (0.88)} & 0.084 {\small (0.025)} & 1.104 {\small (0.074)} & 9.29 {\small (0.37)} & 24 & 1 \\ 
   64 &  3675 & 2.97 {\small (0.61)} & 0.081 {\small (0.019)} & 0.963 {\small (0.028)} & 9.87 {\small (0.30)} & 28 & 8 \\ 
  128 & 14707 & 4.37 {\small (0.51)} & 0.104 {\small (0.010)} & 1.024 {\small (0.023)} & 10.24 {\small (0.41)} & 46 & 60 \\ 
  256 & 58912 & 4.06 {\small (0.27)} & 0.101 {\small (0.004)} & 1.006 {\small (0.021)} & 9.88 {\small (0.38)} & 67 & 429 \\ 
  512 &235730 & 4.02 {\small (0.09)} & 0.098 {\small (0.002)} & 1.008 {\small (0.012)} & 9.61 {\small (0.38)} & 99 & 2985 \\ 
   \multicolumn{8}{c}{{\em Missing Disk (10\%)}} \\ 
   32 &   923 & 4.29 {\small (0.98)} & 0.092 {\small (0.025)} & 1.089 {\small (0.067)} & 9.32 {\small (0.46)} & 20 & 1 \\ 
   64 &  3675 & 2.86 {\small (0.55)} & 0.084 {\small (0.020)} & 0.937 {\small (0.029)} & 9.86 {\small (0.33)} & 40 & 9 \\ 
  128 & 14743 & 4.49 {\small (0.58)} & 0.105 {\small (0.011)} & 1.034 {\small (0.028)} & 10.26 {\small (0.38)} & 74 & 87 \\ 
  256 & 58979 & 4.04 {\small (0.22)} & 0.101 {\small (0.004)} & 1.003 {\small (0.018)} & 9.85 {\small (0.41)} & 130 & 753 \\ 
  512 &235923 & 3.99 {\small (0.09)} & 0.099 {\small (0.003)} & 1.005 {\small (0.013)} & 9.66 {\small (0.42)} & 257 & 7470 \\ 
   \hline
\hline
\end{tabular}
\caption{MCMC results, power exponential covariance with nugget, for different lattice sizes and sampling designs.  
The table shows the lattice size ($n_1=n_2$), number of observations $n$, posterior means (standard deviations) 
for the unknown parameters, average number of PCG iterations per conditional simulation, and the computational 
run time (in minutes) for 2500 MCMC iterations.  The true parameter values are $\sigma^2=4, \lambda= 0.1,
\alpha=1$ and $\mu=10$. Each row corresponds to one simulated dataset. The PCG tolerance is $\epsilon=10^{-5}$.} 
\label{table:MCMC}
\end{table}
%, $c=0.01$ and $\tau^2=0.04$.

Table~\ref{table:MCMC} reports the parameter estimates, number of PCG iterations, 
and computational run times for the MCMC results.   For each lattice size and 
missingness pattern, we report the posterior means and standard deviations for 
the unknown parameters, the average number of PCG iterations per conditional 
simulation, and total run time in minutes.   There are a number of points to note. 
First, the posterior means are fairly close to the true values, with the 95\% 
posterior intervals (not shown) containing the true values for all parameters in 
all examples.   Second, notice that the posterior standard deviation decreases as 
$n$ increases for all parameters except $\mu$.  The latter is expected under 
fixed-domain asymptotics \citep{Stei:99}, since the degree of `learning' about 
the mean is limited by the size of the domain rather than the lattice size.  

Third, the average number of PCG iterations depends on lattice size and missingness 
pattern. In particular, the number of iterations increases with the lattice size.  
For example, for complete lattices, the average number of PCG iterations increases 
from 5 for a $32\times 32$ grid to 38 for a $512 \times 512$ grid.  Similar 
increases occur for the other two sampling designs.  In addition, complete lattice 
designs require fewer PCG iterations than incomplete lattices.  For example, for 
$128 \times 128$ lattices, the complete design requires 13 iterations, missing at 
random requires 46 iterations, and missing disk requires 74 iterations.  Presumably, 
this occurs because the Vecchia preconditioner provides a better approximation to 
the precision matrix for complete lattices than for incomplete lattices.  To 
quantify these relationships, we fit multiple regressions of the number of 
PCG iterations against grid size and sampling design.  Based on the fitted models
(not shown), we conclude that the number of PCG iterations increases like the 
square root of the number of lattice points, i.e., $I(N)=O(N^{1/2})$.

Finally, note that the MCMC run times are roughly 
proportional to the number of PCG iterations.  For the complete lattice design, 2500 
iterations of the MCMC algorithm requires about one minute for a $32 \times 32$ 
lattice, three minutes for $64\times 64$ lattice, 23 minutes for a $128 \times 128$ 
lattice, and about 2.8 hours for a $256 \times 256$ lattice.  The run times for 
$n_1=128$ and $n_1=256$ are impressive, 
considering that no other `exact' Bayesian methods exist for datasets of this size.

\subsubsection{Maximum Likelihood Analysis}

For maximum likelihood analysis, we use the EM algorithm from Section 3.2 with 
modifications to improve computational efficiency.  In particular, the form of 
the likelihood allows us to use profile methods in the M-step.  Under the model 
in Section 4.1, the conditional expectation is
\begin{equation}
Q(\mu,\sigma^2,\bth|\Th^t)
= -\frac{N}{2}\log \sigma^2 -\frac{1}{2}\log |\C(\bth)| 
-\frac{1}{2\sigma^2}E\left[(\Z-\mu\one)'\C(\bth)^{-1}(\Z-\mu\one)|\Z_o,\Th^t\right].
\label{eqn:QQ}
\end{equation}
Using equations (\ref{eqn:Q1})-(\ref{eqn:Q2}) with $\bmu=\mu\one$ and 
$\C=\sigma^2\C(\bth)$, and differentiating with respect to $\mu$ and $\sigma^2$, 
we find that (\ref{eqn:QQ}) is maximized at 
\begin{eqnarray}
     \hat\mu(\bth) &=& \frac{\one'\C(\bth)^{-1}\widetilde\bmu}{\one'\C(\bth)^{-1}\one} 
                    = \frac{\one'\widetilde\bmu}{N},                     \label{eqn:QQmu}\\
\hat\sigma^2(\bth) &=& E\left[S^2(\bth)|\Z_o,\Th^t\right]/N,             \label{eqn:QQsigma}\\
          \hat\bth &=& \arg \max_{\bth} Q_p(\bth|\Th^t),                 \label{eqn:QQth}
\end{eqnarray}
where $\widetilde\bmu=E(\Z|\Z_o,\Th^t)$ is the conditional mean, 
$S^2(\bth)=(\Z-\hat\mu\one)'\C(\bth)^{-1}(\Z-\hat\mu\one)$ is the generalized sum 
of squares, and $Q_p(\bth|\Th^t)$ is the profile function for $\bth$ obtained by 
substituting $\hat\mu(\bth)$ and $\hat\sigma^2(\bth)$ into (\ref{eqn:QQ}).  Note 
that $\hat\mu(\bth)$ does not depend on $\bth$ because $\C(\bth)^{-1}$ is BCCB, 
so $\hat\mu$ can be computed independently of the other parameters.  The profile 
function for $\bth$ is
\begin{equation}
Q_p(\bth|\Th^t) = -\frac{N}{2}\log \hat\sigma^2(\bth) -\frac{1}{2}\log |\C(\bth)|. 
\label{eqn:QQtheta}
\end{equation}
This function is maximized to obtain the estimate $\hat\bth$.  The estimate for 
$\sigma^2$ is then obtained as $\hat\sigma^2(\hat\bth)$ in (\ref{eqn:QQsigma}).  
However, as noted before, the conditional expectation of $S^2(\bth)$ in 
(\ref{eqn:QQsigma}) is computationally intractible for large datasets, so we 
approximate it using Monte Carlo methods.  The Monte Carlo estimate for the 
expected sum of squares is
\begin{eqnarray}
\hat{S}^2(\bth) =\frac{1}{M}\sum_{i=1}^M S^2_i(\bth)
\label{eqn:mc-mle2}
\end{eqnarray}
where $S^2_i(\bth)=(\Z^{(i)}-\hat\mu\one)'\C(\bth)^{-1}(\Z^{(i)}-\hat\mu\one)$, and
$\Z^{(i)} \sim p(\Z\,|\,\Z_o,\Th^t)$, for $i=1,\ldots,M$ are conditional simulations 
of the complete data generated using the current parameter value $\Th^t$.   We then 
substitute $\hat{S}^2(\bth)$ for $E\left[S^2(\bth)|\Z_o,\Th^t\right]$ in 
(\ref{eqn:QQsigma}) and (\ref{eqn:QQtheta}) to obtain an approximate profile 
function, $\widehat{Q}_p(\bth|\Th^t)$.  This function is then maximized to obtain 
the estimates $\hat\bth$ and $\hat\sigma^2 = \hat\sigma^2(\hat\bth)$.  Since 
$\widehat{Q}_p(\bth|\Th^t)$ cannot be maximized analytically, we use numerical 
methods to obtain $\hat\bth$.

We illustrate the maximum likelihood estimation procedure by generating 50
datasets on a $32 \times 32$ square lattice, assuming a complete lattice design
and two incomplete designs (missing at random and missing disk) with three 
missingness probabilities (10\%, 25\%, 50\%).  
We assume an exponential covariance with no microscale variation by fixing 
$\alpha=1$ and $c=0$.   The true values for the unknown parameters are 
$\sigma^2=2$, $\lambda=0.141$ and $\mu=0$.  The exponential model is used in 
this example because it has a closed-form spectral density, which is needed 
for the two competing spectral methods described below.

For the EM algorithm, we use the approach described above with a Monte Carlo sample 
size of $M=400$, using the Vecchia preconditioner and a PCG tolerance of 
$\epsilon=10^{-5}$ as in Section 4.1.1.  For comparison, we also implement two 
approximate maximum likelihood methods: the composite likelihood approach of 
\cite{Vecc:88} and \cite{SteiChiWelt:04}, using prediction sets of size 4 and 
conditioning sets of size 52; and the spectral approximations of \cite{Whit:54} 
and \cite{Fuen:07} for complete and incomplete lattices, respectively.  Note that 
we also considered other approximate likelihood methods (e.g., covariance tapering), 
but the composite likelihood and spectral methods were easier to implement and 
generally gave more accurate results, so the other results are not reported here.

% latex table generated in R 3.0.1 by xtable 1.7-1 package
% Tue Feb 11 10:17:37 2014
\begin{table}[tbp]
\centering
\begin{tabular}{lccccccccccccccc}
  \hline \hline && \multicolumn{4}{c}{$\sigma^2$} && \multicolumn{4}{c}{$\lambda$} 
            && \multicolumn{4}{c}{$\mu$} \\ \cline{3-6} \cline{8-11} \cline{13-16} 
  
            Design && $R^*$ & $R_1$ & $R_2$ & $R_3$ && $R^*$ & $R_1$ & $R_2$ & $R_3$ && $R^*$ & $R_1$ & $R_2$ & $R_3$ \\
  \hline 
    Complete  &  & 450 & 26 &  45 & 387 &  & 35 &  3 &  4 &  33 &  & 550 & 2 &  22 & 237 \\ 
  Random 10\% &  & 446 & 31 &  58 & 596 &  & 47 &  3 &  6 &  92 &  & 545 & 2 &  54 & 231 \\ 
  Random 25\% &  & 450 & 80 &  75 & 626 &  & 50 &  8 &  8 & 127 &  & 556 & 3 &  93 & 220 \\ 
  Random 50\% &  & 457 & 25 & 132 & 552 &  & 49 &  2 & 13 & 147 &  & 558 & 3 & 156 & 227 \\ 
  Disk 10\% &  & 442 & 26 & 268 & 370 &  & 47 &  3 & 25 &  40 &  & 554 & 3 & 393 & 212 \\ 
  Disk 25\% &  & 466 & 24 & 207 & 385 &  & 48 &  2 & 21 &  51 &  & 557 & 3 & 314 & 174 \\ 
  Disk 50\% &  & 491 & 60 & 317 & 351 &  & 47 &  6 & 31 &  65 &  & 554 & 4 & 399 & 163 \\ 
   \hline
\hline
\end{tabular}
\caption{Maximum likelihood estimation, exponential covariance model, 32 $\times$ 32 lattice, 
         different sampling designs, based on 50 simulated datasets.  We report the RMSE ($R^*$) 
         and RMSD ($R_k$) for three competing methods: $R_1 =$ EM algorithm; 
         $R_2 =$ composite likelihood; $R_3 =$ spectral approximation.
         Here, $R^*=(\frac{1}{50}\sum_{r=1}^{50} (\hat\theta_r - \theta)^2)^{1/2}$ and
         $R_k=(\frac{1}{50}\sum_{r=1}^{50} (\hat\theta_{k,r} - \hat\theta_r)^2)^{1/2}$,
         where $\theta$ is the true parameter value, 
         $\hat\theta_r$ is the exact MLE, and
         $\hat\theta_{k,r}$ is the approximate MLE for method $k$.
         True parameter values are $\sigma^2=2$, $\lambda=0.141$, $\mu=0$.
         All values are multiplied by 1000.}
\label{table:RMSD}
\end{table}
%         which are defined for each parameter $\theta \in \{\sigma^2,\lambda,\mu\}$ as 
%         
%         
%         where $\hat\theta_r$ is the MLE for replicate $r$ and $\theta$ is the true parameter.  

Table~\ref{table:RMSD} shows the root mean squared differences (RMSD) for 
each of the estimation methods: EM algorithm, composite likelihood, and spectral
approximation.  For each unknown parameter $\theta \in \{\sigma^2,\lambda,\mu\}$,
and each estimation method $k=1,2,3$, we define 
$$
\mbox{RMSD}_k = 
\left(\frac{1}{50} \sum_{r=1}^{50} (\hat\theta_{k,r} - \hat\theta_{r})^2\right)^{1/2},
$$
where $\hat\theta_{k,r}$ denotes the approximate ML estimate for method $k$, and 
$\hat\theta_{r}$ denotes the exact ML estimate, for each replicate $r=1,2,\ldots,50$.  
The exact MLE can be computed due to the small sample size in this example 
($n\le 1024$).  Note that our definition of RMSD depends on the difference between 
the approximate MLE and the true MLE, rather than the true parameter.  This allows 
us to focus on the approximation error of the estimate (i.e., an RMSD of zero 
corresponds to an estimate with no approximation error.)  For comparison, we also 
compute the root mean squared error for the exact MLE, which is defined as 
$\mbox{RMSE}=(\frac{1}{50}\sum_{r=1}^{50} (\hat\theta_r-\theta)^2)^{1/2}$, 
where $\theta$ is the true parameter value.  For all methods, we use a 
modified Nelder-Mead simplex algorithm for numerical maximization.

Overall, our method outperforms the other two methods for all spatial designs and 
parameters, except one.  First, note that the spectral methods are highly inaccurate, 
with RMSDs that are often 100 times larger than our approach.  The large RMSD for the 
spectral approach is largely due to a strong negative bias for the sill and range 
parameters (not shown).  This bias has been well documented \citep{Guyo:82}, and 
many improvements have been proposed, including data tapering \citep{DalhKuns:87}.  
However, we repeated the analysis using data tapering and the results did not improve 
much.   Second, composite likelihood is more accurate than the spectral approximation, 
but worse than our approach.  Its performance degrades as the proportion of missing 
data increases.  It performs especially poorly for the missing-disk design.  This 
indicates that the local approximation for composite likelihood breaks down near the 
boundaries of large contiguous blocks of missing data.  The composite likelihood 
results could possibly be improved by including some distant neighbors in the 
conditioning set \citep{SteiChiWelt:04}; however, it is not obvious how to 
determine a neighborhood scheme that would work well for all designs.

For the sill and range parameters, $\sigma^2$ and $\lambda$ the RMSDs for our
approach are about 30\% lower for complete lattice, and 50--90\% lower for 
the incomplete lattices.  For the mean parameter $\mu$, the results are 
particularly striking: our estimates are substantially more accurate than the 
other two approaches, with our RMSDs 90--99\% lower than the competing methods.

\subsection{Application to Satellite Sea Surface Temperatures}
\begin{figure}[tbp]
\begin{center}
\includegraphics[width=15cm]{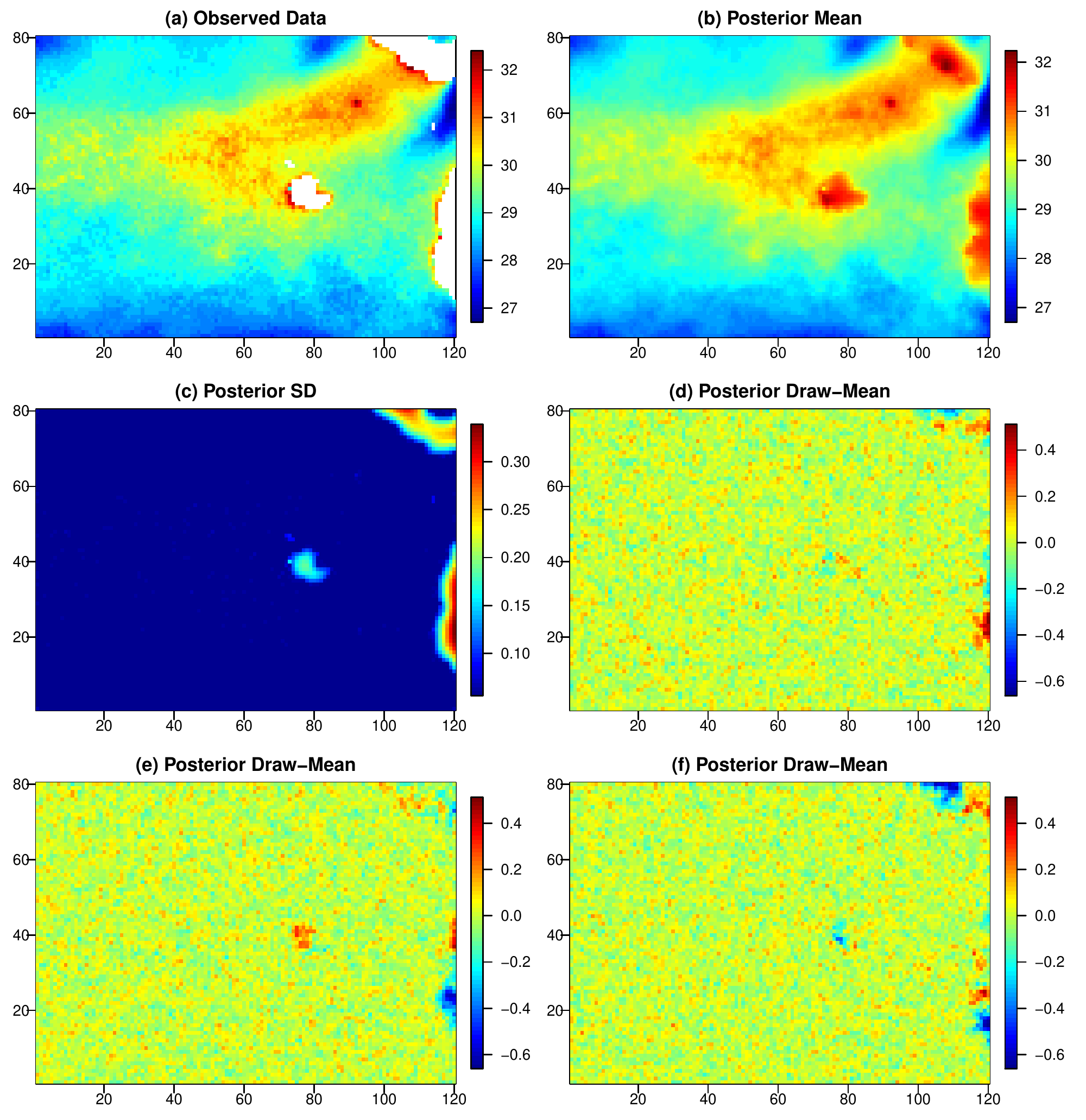}
\caption{TMI satellite data. (a) Observed data $Y(\s)$. 
(b)-(c) Posterior mean and standard deviation for $Z(\s)$. 
(d)-(f) Posterior draws minus the posterior mean for $Z(\s)$.}
%The spatial distance units are pixels.  Each pixel is $25 \times 25$ km.
%Distances can be converted to km by multiplying the axes by 25.}
\label{fig:TMI-maps}
\end{center}
\end{figure}

For a real application, we consider a composite TMI satellite image of sea surface 
temperatures over the Pacific Ocean during the month of March 1998, shown in 
Figure~\ref{fig:TMI-maps}(a).  This dataset was also analyzed by \cite{Fuen:07},
who used spectral methods to fit a stationary Gaussian process to the data 
for maximum likelihood estimation. The observations represent monthly 
averages of sea surface temperatures in the Pacific Ocean on a 120 $\times$ 80 grid; 
the process is undefined over land regions in South and Central America and the 
Galapagos Islands.  The process appears to be somewhat nonstationary, but for 
illustration, we treat it as a stationary process with unknown mean and Mat\'ern 
covariance function with nugget effect.   The percentage of missing data is about 
4\%, and the total number of observations is $n=9203$.

To compare our results with \cite{Fuen:07}, we assume that the observed data
$Y(\s)$ can be written as $Y(\s)=Z(\s)+\varepsilon(\s)$, where $\varepsilon(\s)$ 
is a white noise process with mean zero and variance $\tau^2$, and $Z(\s)$ is a 
stationary, isotropic process with Mat\'ern covariance function \citep{Stei:99}, 
$C(h)=\sigma^2\varphi(h)$, where 
\begin{equation}
\varphi(h) = \frac{1}{2^{\nu-1}\Gamma(\nu)}\left(\frac{h}{\lambda}\right)^{\nu}\mathcal{K}_\nu\left(\frac{h}{\lambda}\right),
\end{equation}
where $\sigma^2$ and $\lambda$ are the sill and range parameters, $\nu>0$ 
is the smoothness parameter, $\mathcal{K}_\nu$ is a modified Bessel function of 
the third kind of order $\nu$ \citep[see][sec. 9]{AbraSteg:64} and $\tau^2$ is 
the nugget effect.  Note that the model in \cite{Fuen:07} also included two 
anisotropy parameters; however, both parameters were found to be insignificant, 
implying that the process is isotropic.  We implement the Bayesian MCMC approach 
described in Sections 3.1 and 4.1.1 to estimate the parameters $\mu, \sigma^2$ 
and $\bth=(\lambda, \nu, c)$ where $c=\tau^2/\sigma^2$ is the noise-to-signal 
ratio.  We assume the following independent prior distributions: 
$\pi(\mu,\sigma^2)\propto \sigma^{-2}$, $\pi(\lambda) \propto 0.5(1+0.5\lambda)^{-2}$, 
$\nu \sim \UU(0,50)$ and $c \sim \UU(0,10)$.

Because of the non-square lattice, our embedding approach requires slight 
modification.  We embed the original $120\times 80$ lattice in a square lattice 
of size $320 \times 320$.  We define the maximum distance in the original domain, 
$\sqrt{120^2+80^2}\approx 144.2$ pixels, to be 1 distance unit. We then apply 
the embedding approach as in Section 4.1 with a radius of $r = 1.5/\sqrt{2} 
\approx 1.06$.  This implies that the minimum size for the embedding lattice 
is $2r\cdot 144.2 \approx 305.7$, which is then rounded up to the next highly 
composite integer of 320.  Note that this choice of radius and embedding 
lattice led to a small number of negative eigenvalues in the sampler.   We 
discarded these parameter draws, under the prior assumption that the 
parameters are constrained to values that yield non-negative definite embeddings.  
The MCMC algorithm described in Section 4.1.1 was run for 65,000 iterations 
after a burn-in period of 35,000.

\begin{figure}[tbp]
\begin{center}
\includegraphics[width=15cm]{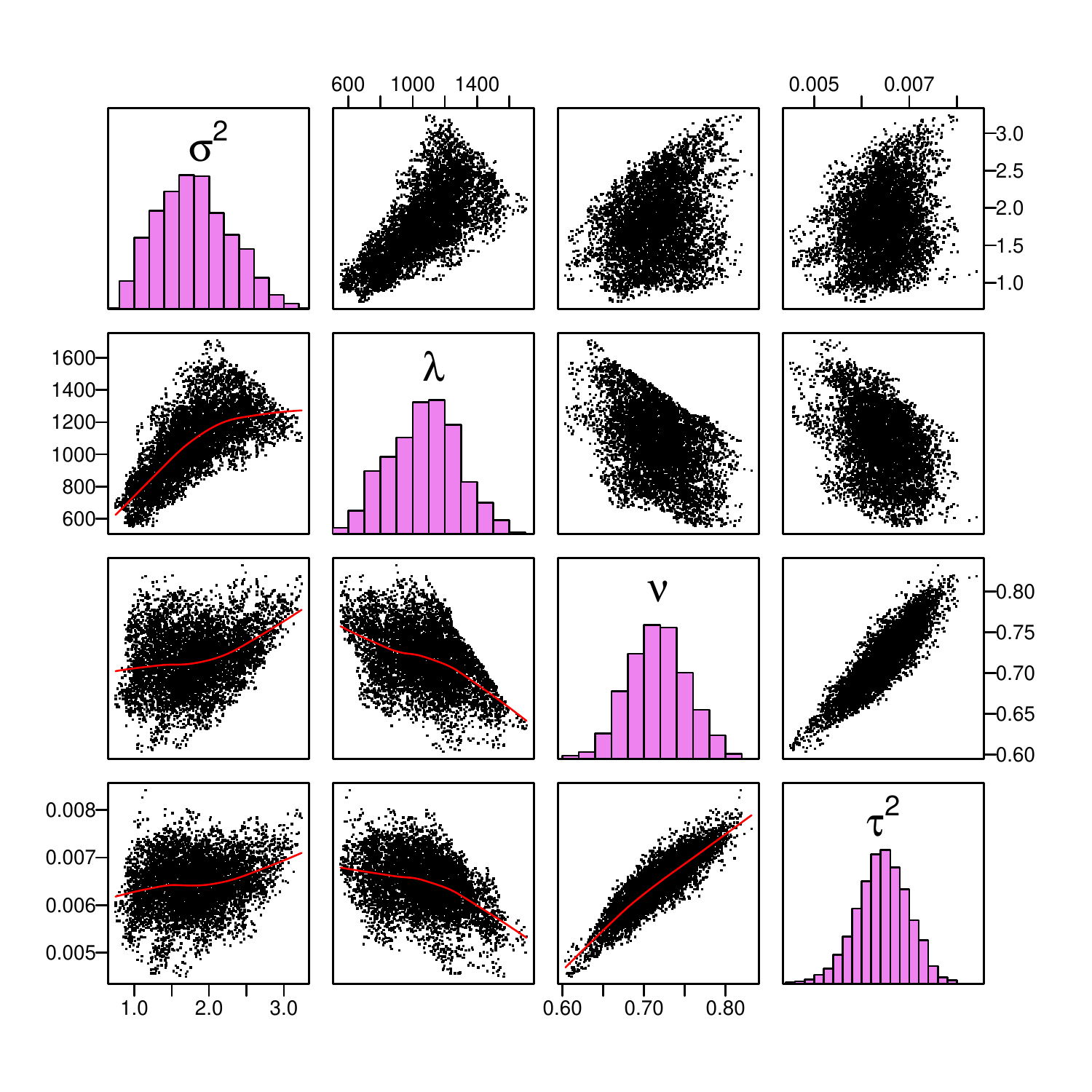}
\caption{TMI satellite data. Posterior histograms and pairs plots for the 
Mat\'ern covariance parameters.  Note that $\lambda$ are in units of km,
which are obtained by multiplying their original distance units by 
(144.2 pixels/distance unit) $\cdot$ (25 km/pixel) = 3605 km/distance unit.}
\label{fig:TMI-pairs-params}
\end{center}
\end{figure}

%(To compare results, we convert $\lambda$ to km by multiplying it by $25 \cdot 144.2 = 3605$ for our method.)

Figure~\ref{fig:TMI-maps} shows the observed data $Y(\s)$, and the posterior mean, 
standard deviation and three posterior draws of the unobserved spatial field, $Z(\s)$.  
Since the nugget effect is relatively small, the posterior mean for the unobserved 
field $Z(\s)$ closely matches the observed data $Y(\s)$ where available, but is 
much smoother.  The posterior standard deviation and draws illustrate the posterior 
uncertainty for $Z(\s)$, which is small except over the land regions.  
Figure~\ref{fig:TMI-pairs-params} shows the posterior histograms and scatterplots 
for the parameters.  Note that the posterior distributions are all unimodal and 
fairly symmetric.  There are dependencies between the parameters, most notably a 
strong positive correlation between the smoothness parameter $\nu$ and the nugget 
effect $\tau^2$.    

Table 3 compares the posterior means and standard deviations for the 
parameters from our MCMC approach with the spectral ML estimates and standard 
errors reported in \cite{Fuen:07}.  For comparison we also calculate the exact
maximum likelihood estimates and their asymptotic standard errors, which are
just barely computable for a sample size of $n=9203$.  
Note that our results differ substantially from the spectral method: our estimates are 
$\hat\sigma^2=1.54$, $\hat\lambda=953$ km, $\hat\nu=0.72$ and $\hat\tau^2=0.006$, 
while those from \cite{Fuen:07} are $\hat\sigma^2=0.57$, $\hat\lambda=312$ km, 
$\hat\nu=1.00$ and $\hat\tau^2=0.001$.  However, the estimates of 
$\sigma^2/\lambda$, which relates to the fine scale variation of the process,
are similar.  The standard errors are also quite different for the two
approaches, in particular for $\sigma^2$ (0.40 vs. 0.02) and $\nu$ (0.04 vs. 1.20).  
On the other hand, our estimates and standard errors agree closely with
the exact ML values for all parameters.  This is quite reassuring.
As a final comparison, we compute the exact loglikelihood values for the three 
sets of estimates.  The loglikelikehood for the spectral estimates is -2077, 
while the MCMC estimates and exact MLEs have nearly identical loglikelihoods 
of 5528, an improvement of 7605 points over the spectral approach.

% latex table generated in R 3.0.1 by xtable 1.7-1 package
% Fri Jan 24 09:12:48 2014
\begin{table}[tbp]
\centering
\begin{tabular}{ccccccc}
  \hline \hline 
Method       & $\sigma^2$& $\lambda$ & $\sigma^2/\lambda$ & $\nu$& $\tau^2$& Loglike \\
  \hline 
Spectral MLE & 0.57 {\small (0.02)} &  312 {\small ( 70)} & .0018 {\small (-------)} & 1.00 {\small (1.20)} & .0010 {\small (.0020)} &  -2077.11\\
%Approx MLE   & 1.37 {\small (0.08)} &  767 {\small ( 87)} & .0018 {\small (-------)} & 0.76 {\small (0.03)} & .0073 {\small (.0005)} &   5526.57 \\
MCMC Bayes   & 1.54 {\small (0.40)} &  953 {\small (215)} & .0016 {\small (.0003)}   & 0.72 {\small (0.04)} & .0066 {\small (.0005)} &   5528.13 \\
Exact MLE    & 1.45 {\small (0.52)} &  911 {\small (312)} & .0016 {\small (-------)} & 0.72 {\small (0.04)} & .0066 {\small (.0006)} &   5528.19 \\
   \hline
\hline
\end{tabular}
\caption{TMI satellite data. Comparison of parameter estimates (standard errors) for the spectral MLE 
results reported in \cite{Fuen:07}, from our Bayesian MCMC approach, and the exact MLEs.
No standard errors are available for $\sigma^2/\lambda$ for the spectral and exact MLEs.
The last column shows the exact log-likelihood values for each set of estimates.}
\label{table:MCMC}
\end{table}

%MCMC-Bayes   & 1.80 {\small (0.49)} & 1070 {\small (213)} & .0017 {\small (.0003)} & 0.72 {\small (0.04)} & .0064 {\small (.0006)} &   5526.03 \\
%Exact MLE   & 1.45 {\small (0.24)} &  910 {\small (189)} & .0016 {\small ()}      & 0.72 {\small (0.04)} & .0065 {\small (.xxxx)} &   5528.18 \\
%Exact MLE   & 1.45 {\small (0.27)} &  910 {\small (186)}    & .0016 {\small ()}      & 0.72 {\small (0.04)} & .0045 {\small (.0009)} &   5528.18 \\
%Exact MLE   & 1.74 {\small ()} &  907 {\small ()} & .0016 {\small ()} & 0.72 {\small ()} & .0045 {\small ()} &   5528.20 \\

\section{Conclusions}
In this paper, we have proposed a new approach for Bayesian and maximum likelihood 
parameter estimation for stationary Gaussian processes observed on a large, 
possibly incomplete, lattice.  We find that the method is feasible for large 
datasets (lattices of up to size $512 \times 512$), it allows for missing data or 
lattices with non-rectangular boundaries, and it provides accurate inference for 
the parameters and missing values.  We propose an MCMC algorithm for Bayesian 
inference and a Monte Carlo EM algorithm for maximum likelihood estimation.   
The method requires the choice of a preconditioner matrix for the PCG algorithm.  
We have developed a number of new preconditioners, and have found that two work 
quite well: the observed block of the complete inverse covariance matrix, and an 
incomplete Cholesky preconditioner based on the composite likelihood methods of 
Vecchia (1988) and Stein et al. (2004).  The latter performs best overall in 
the examples we considered.

The proposed algorithms are widely applicable and conceptually simple, requiring 
iteration between updating the missing data and the parameters.  The main challenges 
in using the method are: (1) selection of a periodic covariance function and an 
embedding lattice that ensures non-negative definiteness of the $\C$ matrix; 
(2) choosing a fast preconditioner that performs well under different parameter 
values, lattice sizes, and missingness patterns; and (3) a need for computational 
code to implement the method.   With regards to the latter, we have developed 
efficient C code for the method, which we plan to make available upon 
publication of this paper.

Possible extensions of this work include estimation of nonstationary processes, 
non-Gaussian processes, multivariate spatial processes, and spatio-temporal 
models.  We have implemented the approach for anisotropic processes; however,
the results are not reported here for space reasons.  We note that another 
estimation approach using circulant embedding is a Gibbs Sampler as proposed 
by \cite{KoziKede:00} for clipped Gaussian fields.  For this approach, the 
unobserved values on the embedding lattice are generated one location at a time.   
This is computationally expensive, requiring $O(N^2)$ operations, where $N$ 
is the number of locations, which is infeasible when $N$ is large.
Our proposed approach based on conditional simulations uses simultaneous 
updating, and thus should provide a more efficient approach to these problems.  
Finally, we also plan to explore the use of parallel 
methods to further improve efficiency of the algorithm.

\bibliographystyle{rss}
\bibliography{gridsim}

\newpage

\section*{Appendix A: BCCB Matrices}%lock-Circulant Matrices}

Let $\C$ be a $N\times N$ block circulant matrix with circulant blocks (BCCB).
Then $\C=\F\La\F^*$, where $\F$ is 
the $2-$dimensional forward Fourier transform (FFT) matrix, $\F^*$ is the inverse FFT
matrix, and $\La= \mbox{diag}(\la)$ is the matrix of eigenvalues, where 
$\la=(\lambda_1,\ldots,\lambda_N)$.  These matrices have a number of computational 
advantages, namely that their eigenvalues, matrix-vector multiplications, and quadratic 
forms can all be computed in $O(N \log N)$ operations using the FFT.
We summarize these operations below.

\begin{my_enumerate}
\item Eigenvalues: $\la=\F\c$ (where $\c$ is the first column of $\C$).
%\item Determinant:  $|\C|=|\La|=\prod_{i=1}^N \lambda_i$.
\item Matrix inverse:  $\C^{-1}=\F\La^{-1}\F^*$.
\item Matrix-vector multiplication: $\C\x=\F\La\F^*\x$. 
%\item Submatrix-vector multiplication: $\C_{oo}\x_{o}=\C\D\x_{o}$ (where $\D$ is a selection matrix)
\item Cholesky-vector multiplication:  $\L\x=\La^{1/2}\F^*\x$ (where $\C=\L\L'$). 
\item Quadratic form:  $\x'\C\x = (\La^{1/2}\F^*\x)'(\La^{1/2}\F^*\x) = \z'\z$.
% (where $\z=\L\x$).
%\item Quadratic form B:  $\x'\C^{-1}\x=\x'\F^*\La^{-1} \F\x=\x'\L'\L\x$. 
\end{my_enumerate}
Multiplications of the form $\F\x$ and $\F^*\x$ are computed using the
$2-$dimensional forward and backward FFT's, respectively, which require $O(N\log N)$ 
operations each.  Matrix-vector multiplications involving the diagonal matrix $\La$ 
are computed by vector-vector multiplications, requiring $O(N)$ calculations.  
Thus, all of the above calculations require $O(N\log N)$ operations.

\section*{Appendix B: Preconditioned Conjugate Gradient}

The preconditioned conjugate gradient (PCG) algorithm is an iterative method to solve 
the system, $\A\x=\b$, where $\A$ is an $n\times n$ symmetric positive-definite 
matrix and $\b \in \mathbb{R}^n$ is known. The PCG method actually solves the equivalent 
system $\M^{-1}\A\x=\M^{-1}\b$, where $\M^{-1}$ is an $n\times n$  preconditioner matrix
chosen such that $\M^{-1}\A$ is better conditioned than $\A$.
The algorithm updates three vectors in $\mathbb{R}^n$: the solution $\x$, the residual $\r$, 
and the search direction $\p$.   The initial solution is defined as $\x_0=\M^{-1}\b$; 
the initial residual and direction are defined as $\r_0=\b-\A\x_0$ and $\p_0=\M^{-1}\r_0$.
The algorithm then proceeds as follows for $k=0,1,2,\ldots$
\begin{eqnarray*}
  \alpha_{k+1} &=& (\r_k'\M^{-1}\r_k)/(\p_k'\A\p_k)\\
      \x_{k+1} &=& \x_k + \alpha_{k+1}\p_k\\ 
      \r_{k+1} &=& \r_k - \alpha_{k+1}\A\p_k\\
   \beta_{k+1} &=& (\r_{k+1}'\M^{-1}\r_{k+1})/(\r_k'\M^{-1}\r_k)\\
      \p_{k+1} &=& \M^{-1}\r_{k+1} + \beta_{k+1}\p_k
\end{eqnarray*}
The algorithm is terminated at the iteration $k=I$ when the residual is less than a specified relative error,
i.e., when $|\r_k|/|\r_0|<\epsilon$, where $\epsilon$ is the specified tolerance.

\section*{Appendix C: Vecchia Preconditioner}
We define the preconditioner $\M^{-1}$ used in the examples.  The goal is to solve the system
$\Si\x=\b$, where $\Si$ is the $n\times n$ covariance matrix for a stationary process on a lattice, and 
$\b$ is a known vector.  We define $\M^{-1}$ as the inverse covariance matrix implied by 
the composite likelihood methods of \cite{Vecc:88} and \cite*{SteiChiWelt:04}.
Let $\Z$ be an $n\times 1$ random vector with distribution $\NN(\Z|\zero,\Si)$.   
For composite likelihood methods \cite*[see][]{VariReidFirt:11}, the observations are ordered 
into $q$ blocks, and the likelihood is written 
as a product of $q$ conditional densities, $p(\A_j\Z|\B_j\Z), j=1,\ldots,q$, where $\A_j$ and 
$\B_j$ are $n_j \times n$ and $m_j \times n$ matrices of zeros and ones defining the prediction 
and conditioning sets for block $j$.  The conditional distributions have the form 
$\NN(\A_j\Z|\K_j\B_j\Z,\V_j)$, where
\begin{eqnarray*}
\K_j &=& \A_j\Si\B_j'(\B_j\Si\B_j')^{-1}\\
\V_j &=& \A_j\Si\A_j'-\A_j\Si\B_j'(\B_j\Si\B_j')^{-1}\B_j\Si\A_j'
\end{eqnarray*}
are matrices of size $n_j \times m_j$ and $n_j \times n_j$, respectively.
Let $\L_j=\A_j-\K_j\B_j$ be the $n_j \times n$  matrix such that 
$\L_j\Z$ are the errors in the regression of $\A_j\Z$ on $\B_j\Z$. 
Then $\L_j\Z \sim \NN(\zero,\V_j)$ independently, and the approximate loglikelihood can 
be written as (ignoring additive constants)
$$
\log p(\Z)  = \sum_{j=1}^q   \left(-\frac{1}{2}\log |\V_j| -\frac{1}{2}  \Z'\L_j'\V_j^{-1}\L_j\Z\right).
$$
This is the loglikelihood for the multivariate normal density $\NN(\Z|\zero,\V)$, where
$$
\V^{-1} \,= \, \sum_{j=1}^q \L_j'\V_j^{-1}\L_j.
$$
This suggests that $\V^{-1} \approx \Si^{-1}$, and therefore we choose $\V^{-1}$ as the 
preconditioner.  This matrix is sparse, and can be fully represented in terms of the 
smaller matrices $\K_j$ and $\V_j$.  Many of these matrices are identical due to the 
stationarity assumption and the lattice domain, so we only need to compute and store
the unique $\K_j$ and $\V_j$.  In concise form, we can write $\V^{-1}=\L'\D\L$, where 
$\D=\mbox{block diag}(\V_1^{-1},\ldots,\V_q^{-1})$ and $\L=[\L_1'~ \L_2' ~ \cdots ~ \L_q']'$ 
are sparse $n\times n$ matrices.  To perform multiplications $\w=\V^{-1}\x$,
we first compute $\u=\D\L\x$, and then $\w=\L'\u$, which can be
computed efficiently by summing over the blocks as follows:
$$
\u=\D\L\x \, = \, \sum_{j=1}^q \V_j^{-1}\L_j\x,
$$
$$
\w=\L'\u \, = \, \sum_{j=1}^q \L_j'\u.
$$

\end{document}